\documentclass[prd,letterpaper,twocolumn,aps,showpacs,nofootinbib,nobibnotes,superscriptaddress,preprintnumbers,floatfix]{revtex4-2}

\usepackage{graphicx,bm,color,braket,amsmath,amssymb,amsfonts}
\graphicspath{{./fig/}{./png/}}

\usepackage{caption}
\usepackage{subcaption}
\usepackage{xcolor}
\usepackage[colorlinks=true,linkcolor=red,urlcolor=blue,citecolor=blue]{hyperref}

\usepackage[normalem]{ulem}
\newcommand{\cMpch}{\,h^{-1}{\rm cMpc}}
\newcommand{\ckpch}{\,h^{-1}{\rm ckpc}}

\newcommand{\cMpchI}{\,h\,{\rm cMpc^{-1}}}
\newcommand{\cMpc}{\,{\rm cMpc}}

\newcommand{\EQ}{\begin{equation}}
\newcommand{\EN}{\end{equation}}
\newcommand{\EQA}{\begin{eqnarray}}
\newcommand{\ENA}{\end{eqnarray}}

{}
{}
{}

\def \pa{\partial}

\newcommand{\pc}{\,{\rm pc}}
\newcommand{\kpc}{\,{\rm kpc}}
\newcommand{\Mpc}{\,{\rm Mpc}}

\newcommand{\aap}{A\&A}
\newcommand{\mnras}{MNRAS}

\begin{document}

\title{Forward cascade of large-scale primordial magnetic fields during structure formation}

\date{\today}

\author{Molly~Abramson }
\email{mabramso@andrew.cmu.edu}
\affiliation{McWilliams Center for Cosmology and Department of Physics, Carnegie Mellon University, Pittsburgh, PA 15213, USA}

\author{Emma~Clarke}
\email{emmaclar@andrew.cmu.edu}
\affiliation{McWilliams Center for Cosmology and Department of Physics, Carnegie Mellon University, Pittsburgh, PA 15213, USA}

\author{Tina~Kahniashvili}
\email{tinatin@andrew.cmu.edu}
\affiliation{McWilliams Center for Cosmology and Department of Physics, Carnegie Mellon University, Pittsburgh, PA 15213, USA}
\affiliation{School of Natural Sciences and Medicine, Ilia State University, 0194 Tbilisi, Georgia}
\affiliation{Abastumani Astrophysical Observatory, Tbilisi, GE-0179, Georgia}

\author{Sayan~Mandal}
\email{sayanm@andrew.cmu.edu}
\thanks{Corresponding author; all authors listed alphabetically.}
\affiliation{McWilliams Center for Cosmology and Department of Physics, Carnegie Mellon University, Pittsburgh, PA 15213, USA}
\affiliation{School of Natural Sciences and Medicine, Ilia State University, 0194 Tbilisi, Georgia}

\author{Salome~Mtchedlidze}
\email{salome.mtchedlidze@unibo.it}
\affiliation{Dipartimento di Fisica e Astronomia, Universit\`{a} di Bologna, Via Gobetti 92/3, 40121, Bologna, Italy}
\affiliation{School of Natural Sciences and Medicine, Ilia State University, 0194 Tbilisi, Georgia}

\author{Jennifer~Schober}
\email{schober@uni-bonn.de}
\affiliation{Argelander-Institut f\"ur Astronomie, Universit\"at Bonn, Auf dem H\"ugel 71, 53121 Bonn, Germany }

\begin{abstract}
The origin of large scale magnetic fields in the Universe is widely thought to be from early Universe processes, like inflation or phase transitions.
These magnetic fields evolve via magnetohydrodynamic processes until the epoch of recombination.
When structures begin to form in the later Universe, the conservation of magnetic flux amplifies the magnetic fields via the adiabatic collapse of gravitationally bound gas clouds hosting the magnetic fields and moves them to smaller scales.
In this work, we have semi-analytically studied this forward cascade effect, considering simple models of gravitational collapse of structures.
We find that this simple model is able to reproduce the general qualitative features of the evolution of the magnetic field spectrum as seen from magnetized cosmological simulations.

\end{abstract}

\maketitle

\section{Introduction}

Magnetic fields are observed throughout the Universe on scales ranging from planets and stars to galaxies and galaxy clusters~\cite{Subramanian:2015lua}. 
The origin of these magnetic fields is an open question in modern astrophysics and cosmology.
It is commonly assumed that the observed cosmic magnetic fields originated from small ``seed” fields that were amplified during structure formation.
Such theories of field generation, or magnetogenesis, can be divided into two major classes: astrophysical (where weak initial seed fields from local sources in galaxies are amplified and transferred to larger scales; or are produced on large scales through charge separation processes; see, e.g., Refs.~\cite{Naoz:2013wla, LangerDurrive18, Kulsrud:2007an}) and cosmological or primordial (where a seed is generated prior to galaxy formation in the pre-recombination Universe on scales that are now large)~\cite{Kandus:2010nw}.

The primordial magnetogenesis scenario is motivated by observations of TeV blazar emission spectra by NASA's Fermi Gamma-ray Space Telescope (see Refs.~\cite{Neronov:2010gir, Tavecchio:2010ja, Essey:2010nd, Taylor:2011bn, Huan:2011kp,Vovk:2011aa, Dolag:2010ni,Takahashi:2011ac, Dermer:2010mm, Finke:2015ona, VERITAS:2017gkr,FermiLAT:2018jdy, Saveliev:2020ynu, AlvesBatista:2020oio, Korochkin:2021vmi, Podlesnyi:2022ydu, MAGIC:2022piy, DaVela:2023maq, Fermi-LAT:2023nas, Dzhatdoev:2023opo, Huang:2023bod, Vovk:2023qfk} and references therein).
That is, the existence of magnetic fields (no weaker than $\sim10^{-16}\,{\rm G}$)
in cosmic voids, i.e.~correlated on Mpc scales, could deflect the electron-positron cascade pairs triggered when multi-TeV photons from blazars interact with extragalactic background light.
This would modify the rechanneling of the TeV emission into the multi-GeV energy range, thereby 
explain
the lack of a multi-GeV bump in some blazar emission spectra.

For this interpretation of the  observed blazar spectra features, at least 60\% of the cosmological volume must be permeated by such fields, a degree of magnetization which is unattainable by magnetized outflows from normal or active galaxies~\cite{Dolag:2010ni, TjemslandEtAl2024}.
Alternative explanations of the Fermi observations have been proposed including uncertainties in the measurements of blazar spectra~\cite{Arlen:2012iy}, plasma instabilities~\cite{Broderick:2018nqf, Yan:2018pca, AlvesBatista:2019ipr, Perry:2021rgv, Perry:2021rgv, Alawashra:2022all, 2023arXiv230301524B,2024arXiv240515390E,Alawashra:2024ndp}, magnetized outflows by normal or active galaxies~\cite{Bondarenko:2021fnn}, return currents induced by cosmic rays escaping into the intergalactic medium from the first galaxies~\cite{Miniati:2010ne, FernandezAlonso:2018rdp}, and collective action of galactic dipole fields \cite{GargEtAl2025}.
Some other alternatives have also been refuted; for example, see~\cite{2025arXiv250514774G}. 
Interestingly, primordial magnetogenesis is supported through the direct detection of a high synchrotron emission flux coming from filaments~\cite{Vernstrom:2021hru} that requires a field strength of the order of $30-60\,{\rm nG}$.
Explanation of these limits~\cite{Hodgson:2021wfy}, as well as the evolution of magnetic fields in cosmic filaments~\cite{Carretti:2022fqk}, remains challenging solely through dynamo amplification and astrophysical mechanisms \cite{Carrettietal2025,Vazzaetal2025}, while cosmological magnetized simulations reveal compatibility with empirical data~\cite{OSullivan:2020pll, Carretti:2022tbj, Pomakov:2022cem, Aramburo-Garcia:2022ywn}, favoring the scenario of tangled primordial magnetic fields (PMFs)~\cite{Vazza:2021vwy,Carrettietal2023,Mtchedlidze:2024kvt}.

In this paper, we focus on the evolution of PMFs during the late stages of the expansion of the Universe, and present a semi-analytical description of the interaction between the magnetic fields and 
large-scale
structures during gravitational collapse.
Previously some of us numerically studied the dynamics of PMFs in the early Universe and the generation of pre-recombination magnetohydrodynamic (MHD) turbulence using the publicly available {\sc Pencil Code}~\cite{Brandenburg:2016odr}.
However, this modeling did not include the effect of the gravitational (density) perturbations -- such a simplification is well justified for the linear stage, but the nonlinearity of density perturbations might substantially change the picture~\cite{Hutschenreuter:2018vkr}.

It is natural to assume that PMFs that were frozen during the dark ages are affected by the physical processes during structure formation. 
Reference~\cite{Mtchedlidze:2021bfy} 
explores some milestones in the evolution of PMFs during structure formation using the cosmological MHD code \textsc{Enzo}~\cite{ENZO:2013hhu}, accounting for different PMF spectra corresponding to different magnetogenesis scenarios and studying the dependence of the field growth on its coherence scale.
It was shown that collapsing structures amplify magnetic fields and move their energy from large to small scales.

In this paper, we theoretically derive the evolution of the magnetic field energy spectrum and show that if the field dynamics is governed by gravitational collapse, then its spectrum should indeed show forward cascading toward smaller scales. 
Our goal is to trace the evolution of the PMF during the formation of large-scale structures in the late Universe.
Specifically, we aim to determine how the energy spectra of PMFs -- characterized by different peak positions and spectral slopes -- evolve and shift under the influence of gravitational collapse.

This paper is arranged as follows.
In Sec.~\ref{secModeling}, we briefly review the statistical modeling of primordial magnetic fields and how their evolution is studied in cosmological simulations.
We introduce the formalism that describes the forward cascade of magnetic energy via collapse of structures in Sec.~\ref{SecForwardCascArg} and present our numerical results in Sec.~\ref{secResults}; finally we present our concluding discussions in Sec.~\ref{secConcl}.
Throughout this article, we work in natural units where 
$\hbar=c=k_\mathrm{B}=1$.
In addition, the permeability of free space is set to unity, $\mu_0 = 1$, expressing all electromagnetic quantities in Lorentz-Heaviside units.

\section{Modeling Cosmological Magnetic Fields }
\label{secModeling}

Primordial magnetogenesis scenarios are typically divided into two categories, depending on their coherence lengths at the moment of PMFs generation:
(a) unlimited by the Hubble horizon length scale and
(b) limited by the Hubble horizon: for example,  inflationary magnetic fields that are generated through quantum-mechanical fluctuations are characterized by arbitrary correlation lengths, since their correlation lengths are increased by the inflationary expansion.
In the opposite case, when the correlation lengths are bounded through the Hubble length (sometimes referred to as {\it causal}) PMFs might arise from primordial turbulence or turbulence associated with violent phase transitions.

There are two cosmological phase transitions of interest in the early Universe: the electroweak (EW) and the quantum chromodynamics (QCD) phase transitions, referred to as EWPT and QCDPT respectively.
In the standard model of particle physics and cosmology, these transitions are smooth crossovers rather than violent first-order PTs; however, much research has been devoted to extend the standard model to make these transitions first-order (see~\cite{Schwarz:2009ii,Rubakov:2017xzr} for discussion).
The high conductivity of primordial plasma ensures a strong interaction of PMFs and plasma motions (which are derived by primordial inhomogeneity); consequently, it leads to the generation of PMFs through induced plasma motions \cite{Quashnock:1988vs}.
Alternatively, magnetic field seeds, generated during the inflationary epoch and afterward that survive the reheating, unavoidably interact with the plasma, leading to the development of turbulence \cite{Kahniashvili:2016bkp}.

\subsection{Statistical Properties}
\label{SecStatProp}

Magnetic fields generated in the early Universe are modeled to have a stochastic\footnote{We do not consider constant spatially homogeneous magnetic fields due to their ``pathological nature" as described in Ref.~\cite{Brandenburg:2020vwp}.
In particular, truly uniform magnetic fields do not demonstrate the familiar magnetohydrodynamic decay properties that, for example, a statistically homogeneous distribution of fields does; see Ref.~\cite{Brandenburg:2020vwp} for more details.} isotropic and statistically homogeneous Gaussian distribution.
The two-point correlation function of these magnetic fields in a real space is defined as
\begin{equation}\label{e2PtCor-1}
\mathcal{B}_{ij}(\mathbf{x},\mathbf{x}+\mathbf{r})\equiv\langle B_i(\mathbf{x})B_j(\mathbf{x}+\mathbf{r})\rangle = \mathcal{B}_{ij}(\mathbf{r}),
\end{equation}
where $B_i$ denotes the $i$-th component of the comoving magnetic field $\mathbf{B}$, $\mathbf{x}$ and $\mathbf{r}$ denote comoving coordinates, and the angle brackets denote ensemble averaging (which we achieve in practice by a spatial average over a comoving volume $V$).
The statistical isotropy (rotational invariance) of the distribution of magnetic fields ensures that the correlation function depends only on $r=|\mathbf{r}|$, i.e., we can write $\mathcal{B}_{ij}(\mathbf{r})=\mathcal{B}_{ij}({r})$.

The three-dimensional power spectrum of magnetic fields $\mathcal{F}_{ij(\mathbf{k})}$, which is the Fourier transform of the real-space correlation function,
\begin{equation}\label{eCorFourier-1}
\mathcal{B}_{ij}(\mathbf{r})=\frac{1}{(2\pi)^3}\int \mathrm{d}^3\mathbf{k}\,e^{-i\mathbf{k}\cdot\mathbf{r}} \mathcal{F}_{ij}(\mathbf{k}),
\end{equation}
can be decomposed into symmetric and helical parts,
\begin{equation}\label{eCorFourier-2}
\frac{{\mathcal F}_{ij}({\bf k}) }{(2\pi)^3}= P_{ij}(\hat{{\bf k}})\frac{E_{\rm M}(k)}{4\pi k^2} + i \varepsilon_{ijl}\,{k_l}\,\frac{H_{\rm M}(k)}{8\pi k^2},
\end{equation}
where $\hat{\mathbf{k}}=\mathbf{k}/k$, $P_{ij}(\hat{{\bf k}})=\delta_{ij}-\hat{k}_i\hat{k}_j$ is the projection operator in momentum space, and $E_{\rm M}(k)$ and $H_{\rm M}(k)$ are respectively the spectral magnetic energy and helicity densities.
In terms of the spectral magnetic energy density, the mean magnetic energy density $\mathcal{E}_{\rm M}=\langle\mathbf{B}^2(\mathbf{x})\rangle/2=\mathcal{B}_{ii}(0)/2$ can be written as
\begin{equation}\label{eEnMean}
    \mathcal{E}_{\rm M}=\int_0^\infty~\mathrm{d}k\,E_{\rm M}(k),
\end{equation}
and we can define an integral length scale $\xi_M$ as
\begin{equation}\label{eInteg}
\xi_{\rm M}=\frac{1}{\mathcal{E}_{\rm M}}\int_0^\infty  ~\mathrm{d}k\,k^{-1}\,E_{\rm M}(k),
\end{equation}
which is the typical scale on which magnetic fields are correlated in the plasma.

Depending on the exact nature of the generation mechanism, magnetic fields can be helical and nonhelical. It has been shown that in the decaying MHD turbulence regime, applicable to the early Universe epoch, the correlation length of both helical and nonhelical fields increases \cite{Brandenburg:2016odr}; 
the field strength and correlation length at the recombination epoch then depend on the magnetic field decay timescale and $B$--$\xi_{\rm M}$ evolutionary trend \cite{Banerjee:2004df,Brandenburg:2017neh,Hosking:2022umv,Brandenburg:2024tyi}. The authors of Ref.~\cite{HoskingSchek2023} argue that the decay timescale is governed by reconnection physics which leads to prolonged decay timescales and therefore to small-scale-correlated (of the order of tens of $\pc$) fields; on the other hand, Ref.~\cite{Brandenburg:2024tyi} finds magnetic field coherence scales reaching hundreds of $\kpc$ at the end of the recombination epoch.\footnote{In this paper we are focused on nonhelical PMFs dynamics during the gravitational collapse, but to show a full picture  of PMFs statistical properties we address helical magnetic fields in Appendix~\ref{SecMagHelRealiz}.}

\subsection{Magnetized cosmological simulations}
\label{SecCosmSims}

Reconstructing the full picture of the evolution of PMFs, from their generation to the current epoch, remains challenging.
MHD codes have been used extensively to study their evolution in the pre-recombination epoch (see~\cite{Vachaspati:2020blt} for a review). 
On the other hand, the post-recombination evolution of PMFs has been explored using cosmological MHD codes; see, e.g., Refs.~\cite{
Marinacci:2015dja,Donnert:2018lbe,Martin-Alvarez:2020enk,Garaldi:2020xos, Vazza:2020phq,Mtchedlidze:2024kvt,Sanati:2024ijt}. 
However, these simulations cannot resolve all scales relevant to the PMF evolution.
For instance, resolving coherence scales down to $\kpc$ or $\pc$ requires extremely high-resolution simulations or zoom-in simulations focused on specific objects, such as galaxies, where large-scale dynamics typically remain underresolved.

Reference~\cite{Mtchedlidze:2021bfy}
instead studied the evolution of different PMFs in a cosmological setup without resolving the turbulent dynamo amplification within galaxy clusters (see, e.g., Ref.~\cite{Vazza:2020phq} where the subgrid dynamo model has been used to account for the turbulent amplification of PMFs in galaxy clusters).
Nevertheless, such simulations can very well capture global properties of PMFs and their evolution on large scales during the formation of large-scale structure. 

In Ref.~\cite{Mtchedlidze:2021bfy}, some of us used the cosmological code \texttt{Enzo}~\cite{ENZO:2013hhu} to study the evolution of PMFs in a $(67.7 h^{-1}\cMpc)^3$ comoving volume (``c'' referring to comoving units) employing $512^3$ grid cells and $512^3$ dark matter particles, from redshift $z=50$ to $z=0$; this corresponds to a $132\,\ckpch$ spatial resolution and a dark matter particle mass of $m_{\text{DM}} = 2.53\times 10^{8} M_{\odot}$.
We assumed $\Lambda$CDM cosmology with parameters $h=0.674$, $\Omega_m=0.315$, $\Omega_b=0.0493$, $\Omega_{\Lambda}=0.685$, and $\sigma_8=0.807$ \citep{Planck2018} and focused on adiabatic physics, neglecting gas cooling, chemical evolution, star formation, and feedback from active galactic nuclei.
For a more detailed description of the simulation setup, numerical methods, and resolution studies, we refer the reader to Ref.~\cite{Mtchedlidze:2021bfy}, while in this section we briefly summarize the main findings of that study.

\begin{figure*}[t]
    \centering
    \begin{subfigure}[b]{0.45\textwidth}
	\includegraphics[width=\textwidth]{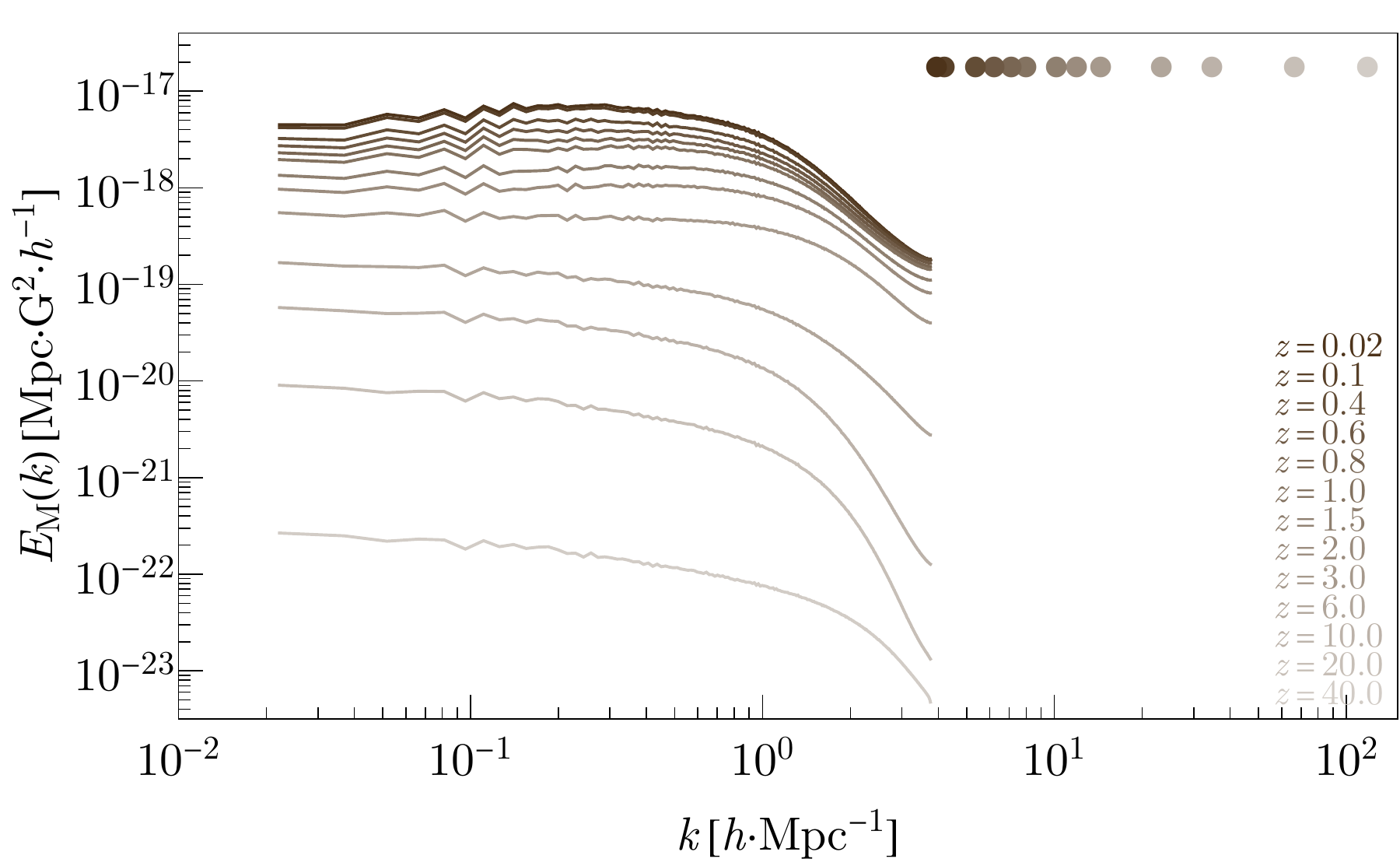}
	\caption{Uniform spectrum.}
	\label{FigSI3Da}
    \end{subfigure}
    ~ 
    \begin{subfigure}[b]{0.45\textwidth}
	\includegraphics[width=\textwidth]{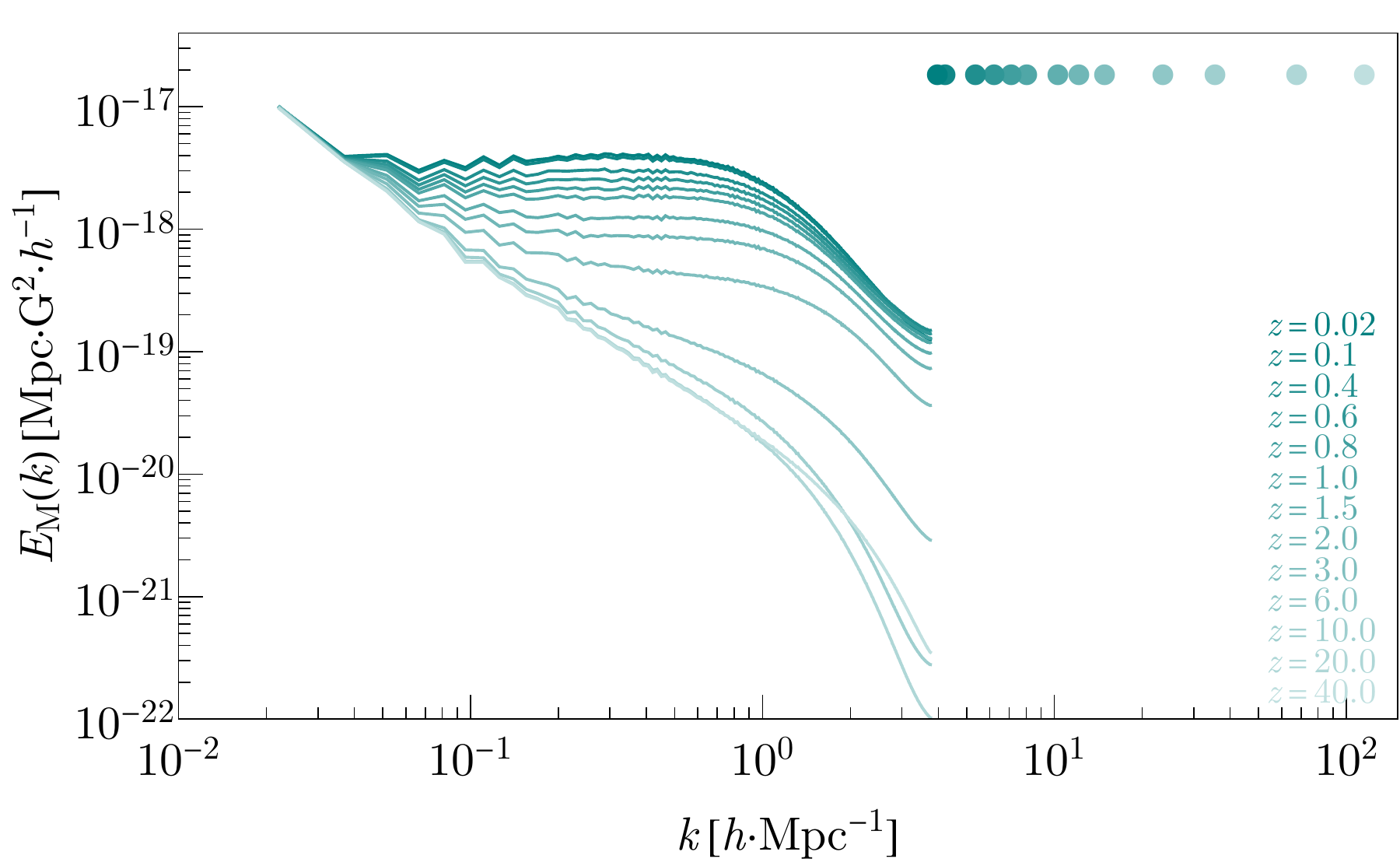}
	\caption{Scale-invariant spectrum.}
	\label{FigSI3Db}
    \end{subfigure}
    \\
    \begin{subfigure}[b]{0.45\textwidth}
	\includegraphics[width=\textwidth]{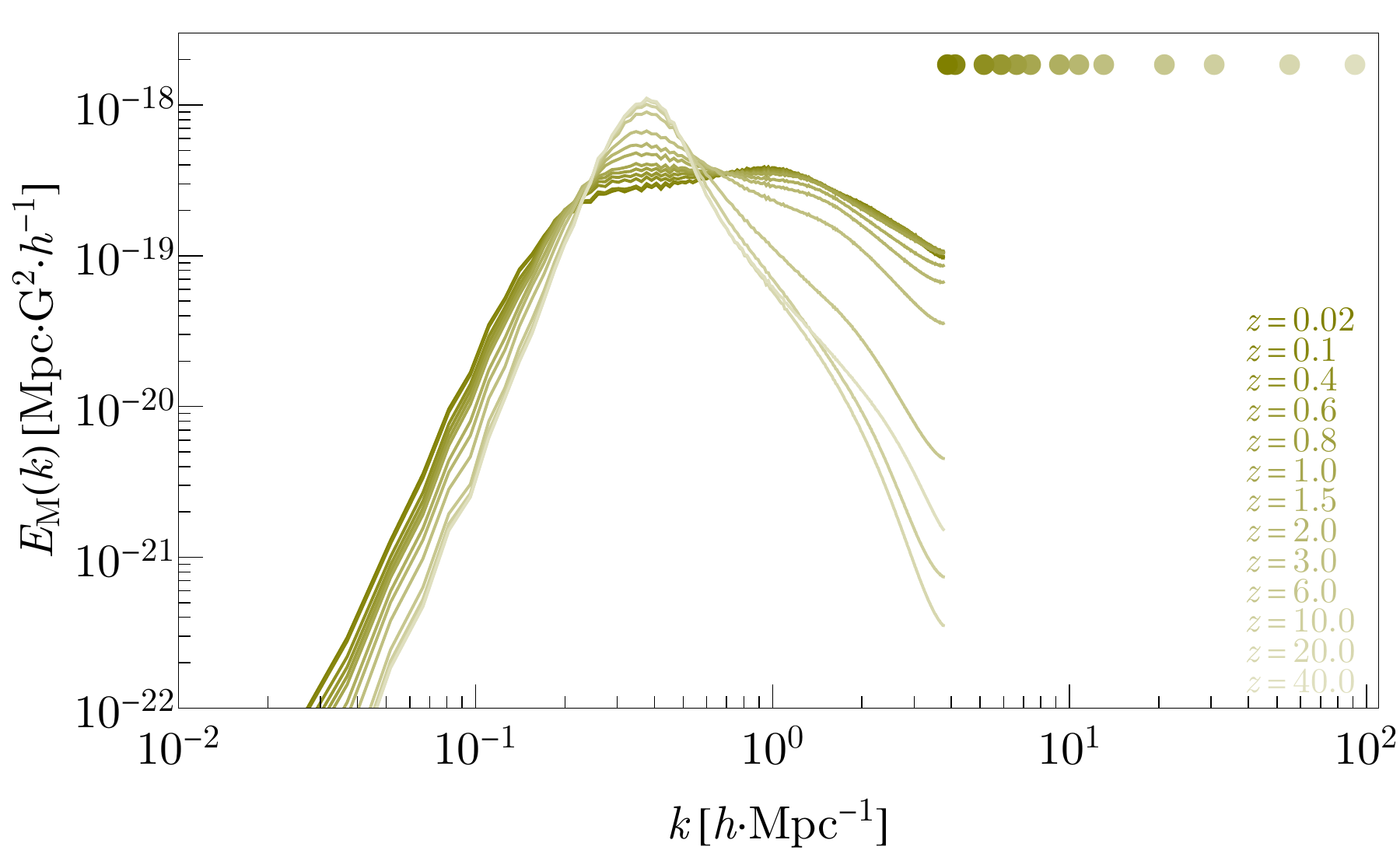}
	\caption{Helical spectrum.}
	\label{FigHel3Dc}
    \end{subfigure}
    ~
    \begin{subfigure}[b]{0.45\textwidth}
	\includegraphics[width=\textwidth]{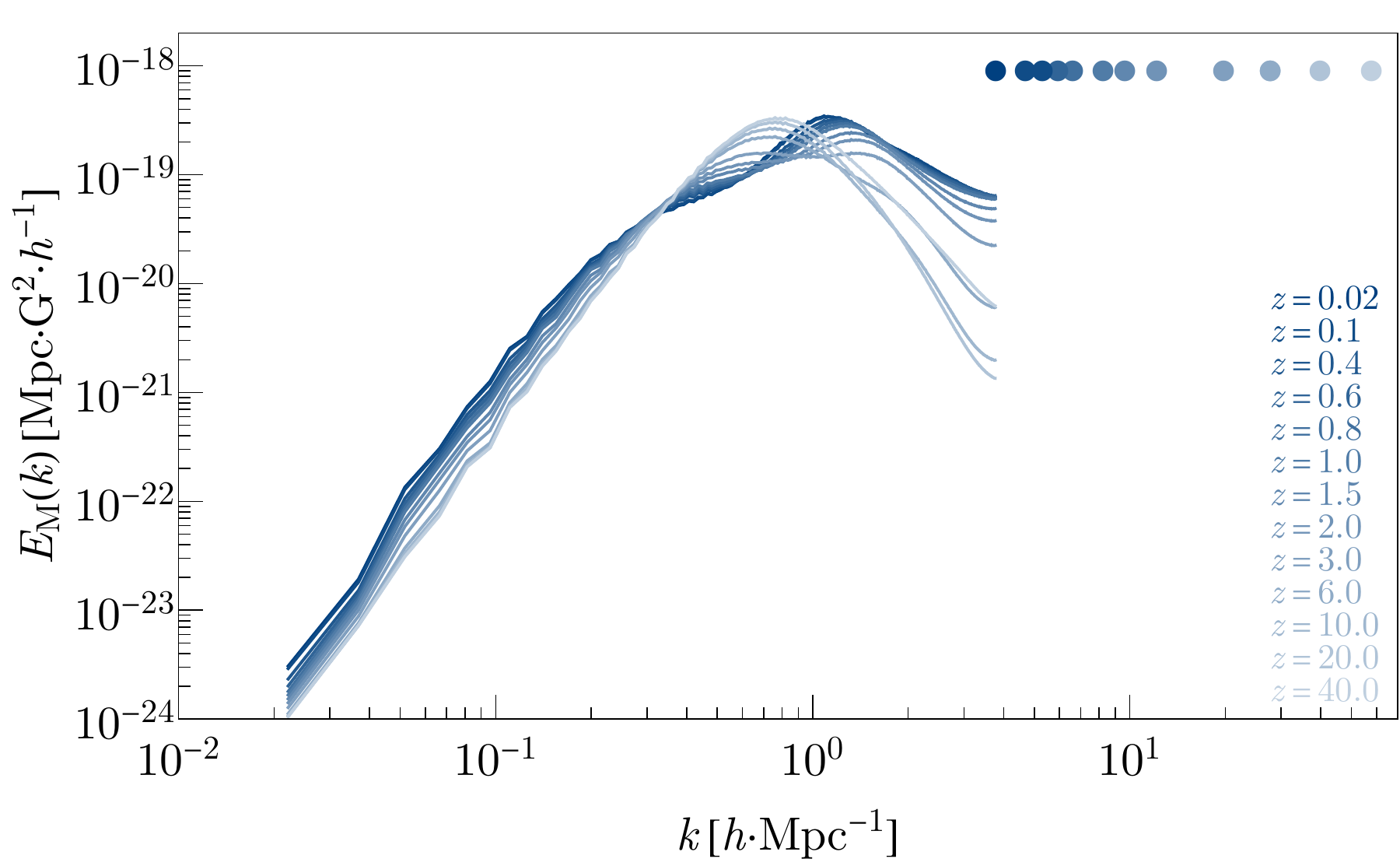}
	\caption{Nonhelical spectrum.}
	\label{FigHel3Dd}
    \end{subfigure}
    \caption{The evolution of the magnetic energy spectra $E_{\rm M}(k)$ with redshift $z$ as affected by structure formation, calculated by numerical simulations in Ref.~\cite{Mtchedlidze:2021bfy}, for different initial spectra for seed PMFs -- uniform, scale-invariant, helical, and nonhelical. The filled circles indicate the Jeans wavenumber at the different redshifts given by $k_\mathrm{J}(z)=a(z)\sqrt{4\pi G \bar{\rho}(z)}/c_\mathrm{s}$.}
    \label{Fig3D}
\end{figure*}

In  Fig.~\ref{Fig3D}, we show the evolution of the PMF energy spectrum $E_{\rm M}(k)$ from Ref.~\cite{Mtchedlidze:2021bfy} for four different PMF models:
(a) a spatially homogeneous model, i.e., a constant magnetic field,
(b) a stochastic magnetic field initially (at $z=50$) characterized by a $k^{-5/3}$ spectra,
(c) a stochastic helical field with a $\sim 2 \cMpch$ correlation length, and
(d) a stochastic field with a coherence scale of $\sim 1 \cMpch$.
We refer to the latter two models respectively as helical and nonhelical, to be consistent with the labelling in Ref.~\cite{Mtchedlidze:2021bfy}; we note, however, that in this work we do not study the effects from helicity.

\begin{figure}[t]
    \includegraphics[width=0.9\columnwidth]{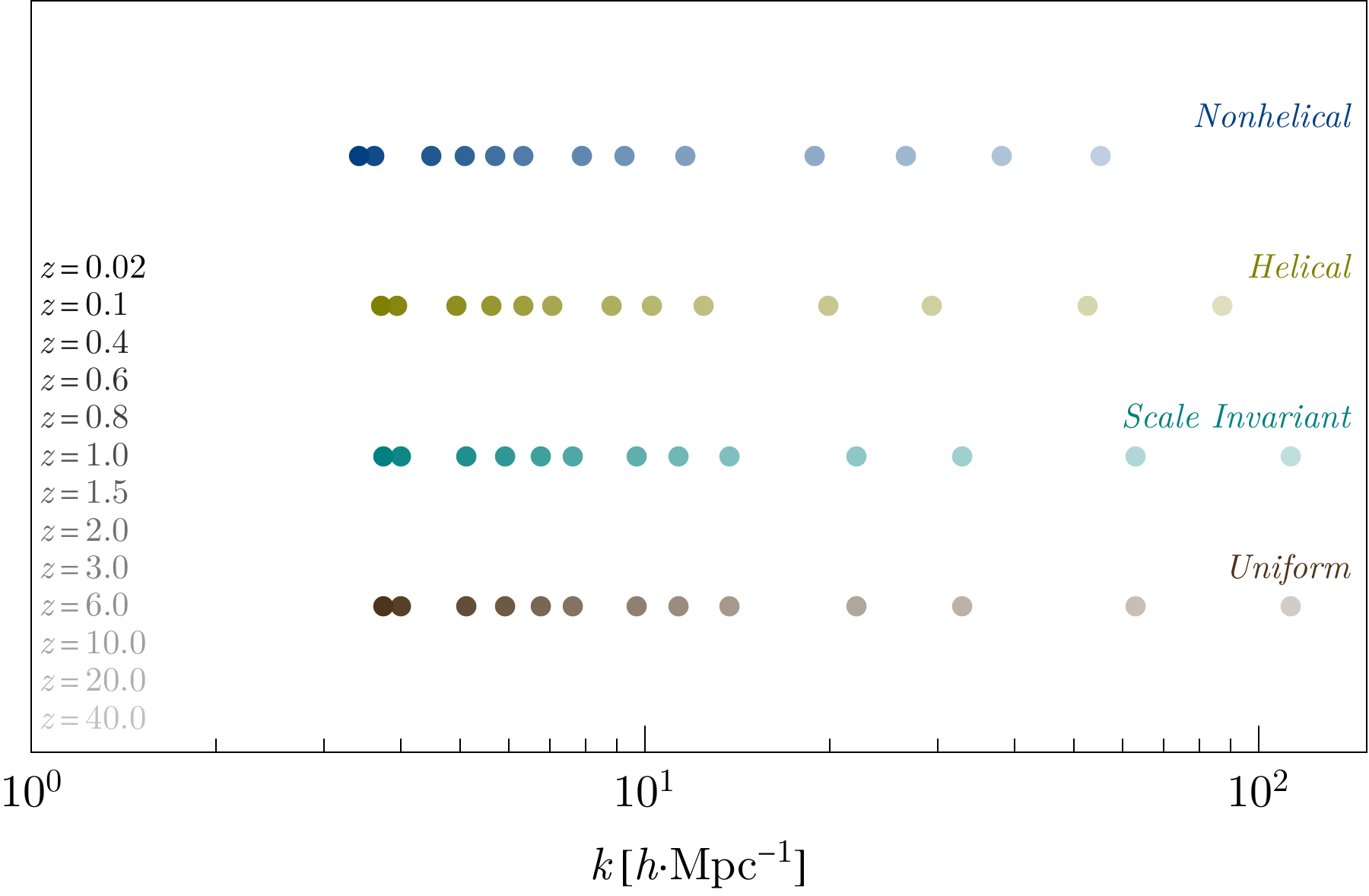}
    \caption{Evolution of the Jeans wavenumber $k_\mathrm{J}(z)=a(z)\sqrt{4\pi G \bar{\rho}(z)}/c_\mathrm{s}$ with redshift $z$ for all four cases in Fig.~\ref{Fig3D}.}
    \label{FigJeansScales}
\end{figure}

From Fig.~\ref{Fig3D}, we see that the spectra of different PMF models evolve in distinguishable ways.
The uniform model exhibits homogeneous growth at small wavenumbers from the first redshifts; the model with initial $k^{-5/3}$ spectra shows pronounced growth only from $z=10$ (the onset of structure formation).
In these two cases, the magnetic field growth is larger on large scales (compared to the growth observed for cases (c) and (d)) since these models have the most of the energy contained on large scales.
Finally, for models (c) and (d), which have a characteristic peak within the simulation box, we see (i) moderate growth on large scales, (ii) a decay of the magnetic energy at the initial peak scale, and then (iii) a shift of magnetic energy toward smaller scales (at $\sim 1 \Mpc$ scales where galaxy clusters form).
In Fig.~\ref{FigJeansScales}, we show the comoving Jeans wavenumber $k_\mathrm{J}(z)=a(z)\sqrt{4\pi G \bar{\rho}(z)}/c_\mathrm{s}$ as a function redshift $z$ for all four cases in Fig.~\ref{Fig3D}; here, $a(z)$ is the scale factor, $c_\mathrm{s}$ is the speed of sound, and $\bar{\rho}$ is the average \textit{physical} density of matter (i.e., of gas and dark matter combined).

In Appendix~\ref{App:ResEffect}, however, we also show that this shift of the characteristic peak in our simulations is resolution-dependent; this is the case, in general, for the amplification of magnetic fields (see also Refs.~\citep{Vazza:2014jga, Steinwandel:2021exs}) since as resolution increases, turbulent amplification also contributes to the growth of magnetic fields on smaller scales and, therefore, interplay between the adiabatic and turbulent growth of magnetic fields becomes more complex.

\section{Magnetic Forward Cascade from Adiabatic Collapse}
\label{SecForwardCascArg}

\subsection{Adiabatic Collapse and Flux Conservation}
\label{secFluxCon}

After recombination, the physical magnetic fields dilute with redshift $z$ as $\mathbf{B}\propto (1+z)^2$, keeping the comoving magnetic field strength constant.
This is a consequence of the conservation of magnetic flux. 
Once (nonlinear) cosmic structure formation sets in, the magnetic energy spectrum is affected by (i) flux freezing during gravitational collapse and (ii) nonlinear physics governing the injection and dissipation of energy \cite{Kahniashvili:2012dy, Sanati:2024ijt, Subramanian:2015lua, Durrer:2013pga}.
If we consider comoving magnetic fields to be affected only by the adiabatic gravitational collapse of structures, we see that the magnetic fields are amplified as the structures shrink in size.
In particular, if a structure of typical size $L_{\rm{I}}$ hosting a magnetic field of typical strength $B_{\rm{I}}$ at redshift $z_{\rm{I}}$ collapses to size $L_{\rm{F}}$ at a later redshift $z_{\rm{F}}$, the strength of the magnetic fields increases to $B_{\rm{F}}=B_{\rm{I}}\left(L_{\rm{I}}/L_{\rm{F}}\right)^2$.
This should lead to a forward cascade of magnetic energy to smaller length scales, i.e., towards larger wavenumbers, 
which potentially counters the inverse cascade of the energy of helical magnetic fields during their MHD evolution which occurred before recombination.

We can get an estimate of this forward cascade as follows.
Let us choose the function $Q(k)$ to denote how structures collapse adiabatically, i.e.,~structures of typical wavenumber $k$ collapse to wavenumbers $Q(k)$.
We assume that $Q(k)\geq k\,\forall\,k$ to ensure that there is no dilation at any scale and that it is monotonically increasing with $k$ to ensure that $Q^{-1}(k)$ exists uniquely\footnote{Note here that $Q(k)$ represents a wavenumber, and so does $Q^{-1}(k)$; the latter \textit{does not} indicate $1/Q(k)$.}.

Let $\mathrm{d}\mathcal{E}_{\rm M}(k)=E_{\rm M}(k)\,\mathrm{d}k$ be the magnetic energy density contained between the scales $k$ and $k+\mathrm{d}k$.
After collapse, this energy density moves to a scale $Q(k)$, and is replaced by the energy density $\mathrm{d}\mathcal{E}_{\rm M}(Q^{-1}(k))=E_{\rm M}\left(Q^{-1}(k)\right)\,\mathrm{d}k$ previously between the scales $Q^{-1}(k)$ and $Q^{-1}(k)+\mathrm{d}k$. 
Due to amplification of the magnetic field, this energy density is amplified by a factor of $\left(k/Q^{-1}(k)\right)^4$. The new energy density $\mathrm{d}\mathcal{E}^{\rm New}_{\rm M}(k)$
on the scale $k$ is then given by
\begin{equation}\label{eNewEnDenK}
    \begin{aligned}
        \mathrm{d}\mathcal{E}^{\rm New}_{\rm M}(k)&=\left(\frac{k}{Q^{-1}(k)}\right)^4\mathrm{d}\mathcal{E}_{\rm M}(Q^{-1}(k))\\
        &=\left(\frac{k}{Q^{-1}(k)}\right)^4E_{\rm M}\left(Q^{-1}(k)\right)\,\mathrm{d}k.
    \end{aligned}
\end{equation}
In terms of the spectral energy density $E^{\rm New}_{\rm M}(k)$ after the collapse, we can write the LHS of Eq.~\eqref{eNewEnDenK} as
\begin{equation}\label{eNewEnDenK-2}
	\mathrm{d}\mathcal{E}^{\rm New}_{\rm M}(k)=E^{\rm New}_{\rm M}(k)\,\mathrm{d}k.
\end{equation}
Comparing Eqs.~\eqref{eNewEnDenK} and \eqref{eNewEnDenK-2}, we can write
\begin{equation}\label{eNewEnDenK-3} 
    E^{\rm New}_{\rm M}(k)=\left(\frac{k}{Q^{-1}(k)}\right)^4 E_{\rm M}\left(Q^{-1}(k)\right).
\end{equation}

In the following subsection, we derive the evolution equation for the forward cascade of the magnetic field energy spectrum $E_{\rm M}(k,t)$.

\subsection{The Forward Cascade Equation}
\label{SecForwardCascEvolEq}

We start with the simplifying  assumption that all structures at scale $k$ at time $t$ collapse at a rate $\zeta(k,t)$.
This means that a structure on scale $k$ at time $t$ becomes a structure on scale $k+\zeta(k,t)\,\mathrm{d}t$ at time $t+\mathrm{d}t$.
In reality, different structures at the same scale may collapse at different rates depending on how they differ by their mass, density profile, thermodynamic properties, etc.; in Appendix~\ref{SecForwardCascEvolEqMassDep}, we generalize this discussion slightly by considering a mass dependence of the collapse rate.

Structures on scale $k$ at time $t+\mathrm{d}t$ were on scale $k'$ at time $t$, which is related to $k$ as $k'+\zeta\left(k',t\right)\,\mathrm{d}t=k$.
This gives
\begin{equation}\label{eScaleComp-1}
	\begin{aligned}
	k'&=k-\zeta\left(k',t\right)\,\mathrm{d}t\\
	&\approx k-\zeta(k,t)\,\mathrm{d}t,
	\end{aligned}
\end{equation}
where the approximation in the second step induces an error of $\mathcal{O}(\mathrm{d}t^2)$ that we ignore.
Using the result in Eq.~\eqref{eNewEnDenK-3} we write the magnetic energy density spectrum at time $t+\mathrm{d}t$ as
\begin{equation}\label{eNewEnDenKInf}
    \begin{aligned}
	E_{\rm M}(k,t+\mathrm{d}t)&=\left[\frac{k}{k-\zeta(k,t)\,\mathrm{d}t}\right]^4 E_{\rm M}\left(k-\zeta(k,t)\,\mathrm{d}t,t\right)\\
	&=\left[1-\frac{\zeta(k,t)\,\mathrm{d}t}{k}\right]^{-4} E_{\rm M}\left(k-\zeta(k,t)\,\mathrm{d}t,t\right).
	\end{aligned}
\end{equation}
With appropriate binomial and Taylor expansions, and keeping terms up to $\mathcal{O}(\mathrm{d}t)$, we can write
\begin{equation}\label{eNewEnDenKInf-2}
\begin{aligned}
    E_{\rm M}(k,t)+\frac{\pa E_{\rm M}(k,t)}{\pa t}\,\mathrm{d}t&=\left[1+\frac{4\zeta(k,t)\,\mathrm{d}t}{k}\right]\\
    &\times\left[E_{\rm M}(k,t)-\frac{\pa E_{\rm M}(k,t)}{\pa k}\zeta(k,t)\,\mathrm{d}t\right]\\
    &=E_{\rm M}(k,t)\\
    &+\zeta(k,t)\left[\frac{4}{k}-\frac{\pa}{\pa k}\right]E_{\rm M}(k,t)\,\mathrm{d}t\\
    &+\mathcal{O}(\mathrm{d}t^2).
\end{aligned}
    \end{equation}
Simplifying the above equation, we can express the evolution of the spectral magnetic energy density as the \textit{forward cascade equation}
\begin{equation}\label{eNewEnDenKInf-3}
    \frac{\pa E_{\rm M}(k,t)}{\pa t}+\zeta(k,t)\left[\frac{\pa}{\pa k}-\frac{4}{k}\right]E_{\rm M}(k,t)=0.
\end{equation}

\section{Numerical Solutions}
\label{secResults}

\begin{figure}[t]
    \includegraphics[width=\columnwidth]{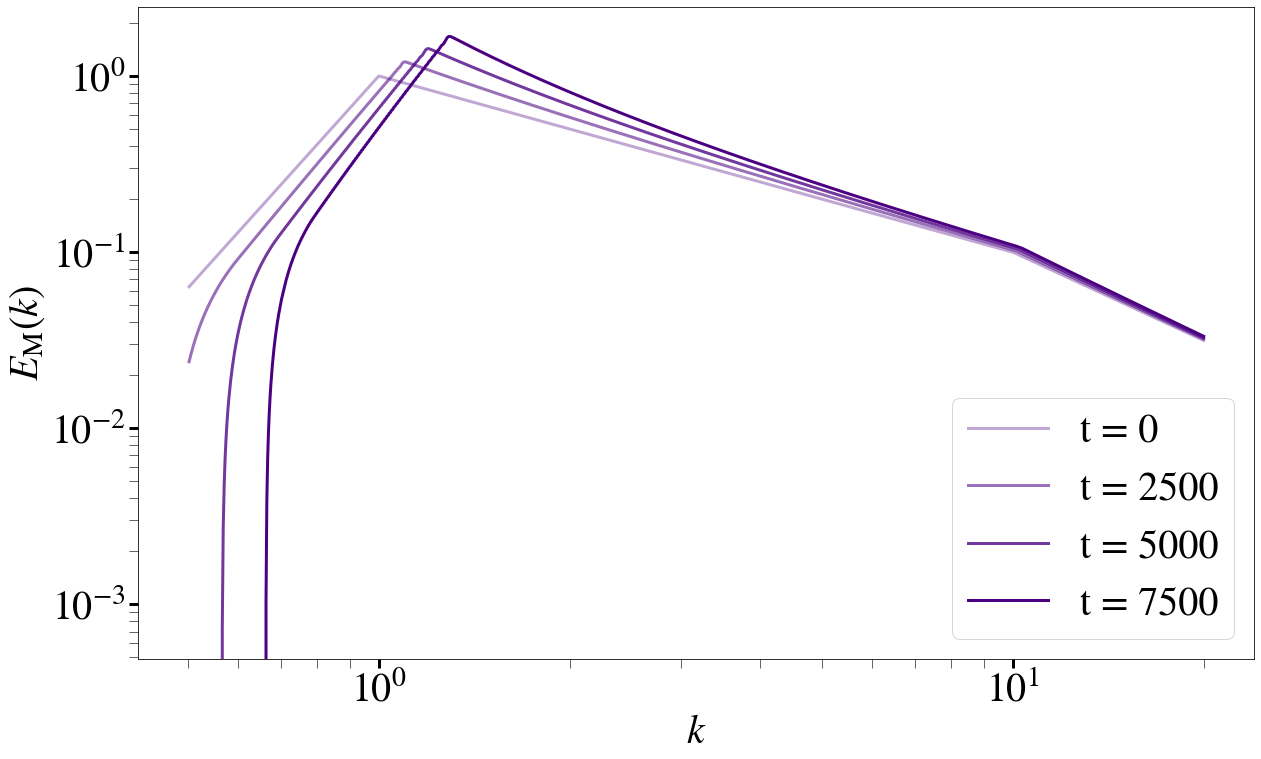}
    \caption{Evolution of the magnetic energy density $E_{\rm M}(k)$ according to Eq.~\eqref{eNewEnDenKInf-3} for a fixed collapse rate, $\zeta_0 = 0.01$. The time $t$ is measured in units of timesteps.
   }
    \label{FigConstRate}
\end{figure}

In order to numerically solve Eq.~\eqref{eNewEnDenKInf-3}, we use the Finite Difference Method which parameterizes partial differentials as finite differentials utilizing a grid representing the solution at some time $t$ and some wavenumber $k$, and solving for times $t+1$ and wavenumbers $k+1$ based on previous temporal and spatial solutions. 
We construct a grid of dimensions $(T,K)$, where $\partial E_{\rm M}/ \partial k \rightarrow (E_{\mathrm{M},t}^{k+1}-E_{\mathrm{M},t}^{k})/\mathrm{d}k$, and $\partial E_{\rm M}/\partial t\rightarrow (E_{t+1}^{k}-E_t^{k})/\mathrm{d}k$;
here, $\mathrm{d}k$ and $\mathrm{d}t$ are the respective step sizes of the grid. 
For the left and right boundaries of the grid, the respective update conditions are
\begin{equation}\label{eLeftUpdate}
    E_{\mathrm{M},t+1}^k = E_{\mathrm{M},t}^k -  \left(\frac{E_{\mathrm{M},t}^{k+1}-E_{\mathrm{M}, t}^{k}}{\mathrm{d}k} - \frac{4}{k} E_{\mathrm{M},t}^k\right)  \zeta_0 \mathrm{d}t ,
\end{equation}
and
\begin{equation}\label{eRightUpdate}
    E_{\mathrm{M},t+1}^k = E_{\mathrm{M},t}^k -  \left(\frac{E_{\mathrm{M},t}^{k}-E_{\mathrm{M},t}^{k-1}}{\mathrm{d}k} - \frac{4}{k} E_{\mathrm{M},t}^k\right)  \zeta_0 \mathrm{d}t  .
\end{equation}
Otherwise, the general update condition is
\begin{equation}\label{eGenUpdate}
    E_{\mathrm{M},t+1}^k = E_{\mathrm{M},t}^k -  \left(\frac{E_{\mathrm{M},t}^{k+1}-E_t^{\mathrm{M},k-1}}{2 ~\mathrm{d}k} - \frac{4}{k}  E_{\mathrm{M},t}^k\right) \zeta_0 \mathrm{d}t ,
\end{equation} 
where the factor of $2$ in the denominator of the partial derivative of $E_{\rm M}$ with respect to $k$ accounts for the larger step size taken in this regime compared to that taken in the boundaries.
For the results demonstrated in the subsequent figures in this section, we have chosen $\mathrm{d}k = 0.002$  and $\mathrm{d}t = 0.002$.

The initial energy spectrum follows a $\propto k^4$ power law from $k =0$ to $k = 1$. From $k = 1$ to $k = 10$, the energy spectrum follows a $\propto k^{-1}$ power law; and for $k > 10$, the energy spectrum follows a $\propto k^{-5/3}$ power law.
In Fig.~\ref{FigConstRate}, we show the evolution of the spectrum $E_{\rm M}(k)$, according to Eq.~\eqref{eNewEnDenKInf-3} for a constant collapse rate $\zeta=\zeta_0$ and the initial conditions above.
We see a clear indication of the forward collapse of the magnetic energy with the collapse of structures.

\begin{figure}[!t]
    \includegraphics[width=\columnwidth]{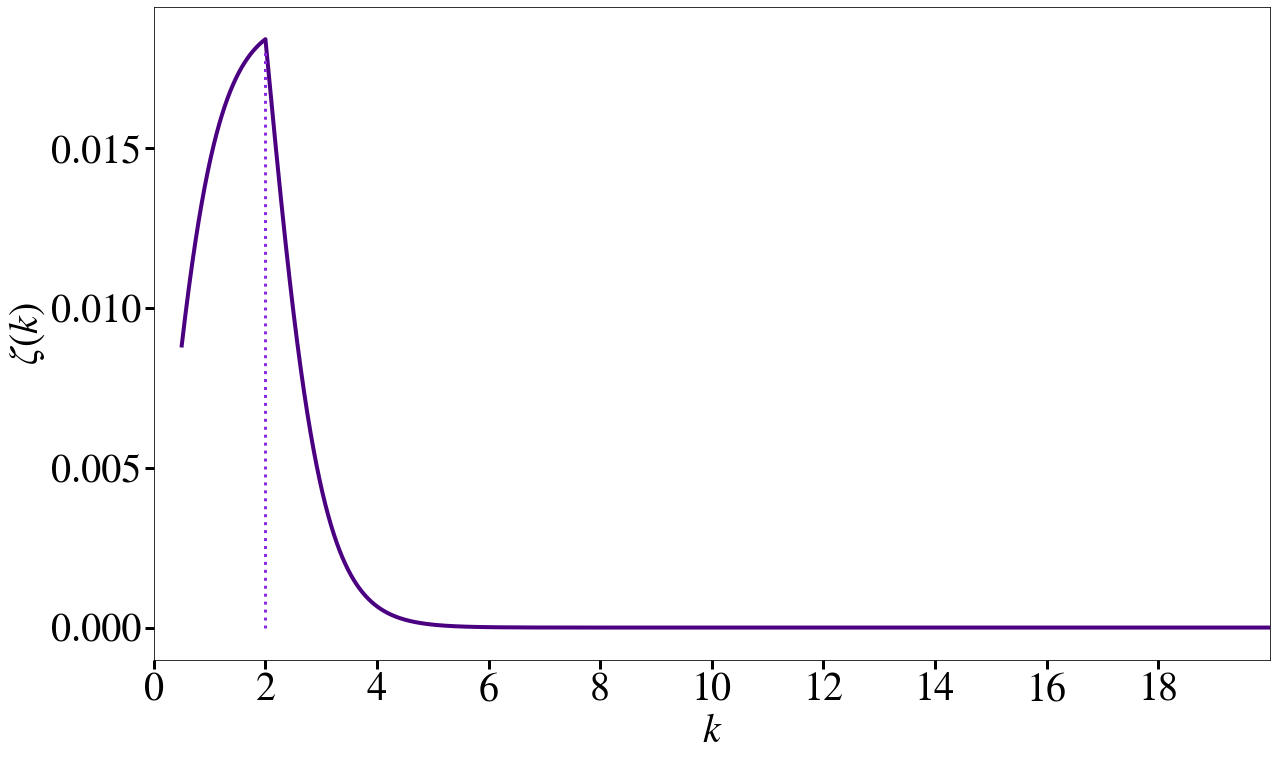}
    \caption{The collapse rate $\zeta(k)$, as defined by Eqs.~\eqref{eJeans} and~\eqref{eZetaSmallK}, plotted as a function of $k$, with $k_{\mathrm{J}} = 2$, as indicated by the vertical line.
    }
    \label{FigZetaPlot}
\end{figure}

However, this constant collapse rate implies that all structures collapse at the same rate at any given time, which is unphysical and does not realistically model the formation of structures.
To have a collapse at a given Jeans wavenumber $k_\mathrm{J}$, the collapse rate is defined from $k=k_\mathrm{J}$ to $k\rightarrow\infty$ to be
\begin{equation}\label{eJeans}
    \zeta(k)=\zeta_0\left[1-\tanh\left\{a(k-k_J)\right\}\right].
\end{equation}
Numerically for small $k$, to ensure that the $k= 0$ to $k = 1$ regime has roughly a power spectrum of $k^4$, changing the evolution of $\zeta_0$ to become a function of $k$ can resolve this issue.
We define $\zeta(k)$ for the range $k\in[0,k_\mathrm{J}]$ to be
\begin{equation}\label{eZetaSmallK}
    \zeta(k)= \frac{\pi \zeta_0}{2} \arctan(k)
\end{equation} 
The resulting collapse rate is shown in Fig.~\ref{FigZetaPlot}.

\begin{figure*}[!t]
    \begin{subfigure}[b]{0.49\textwidth}
	\includegraphics[width=\textwidth]{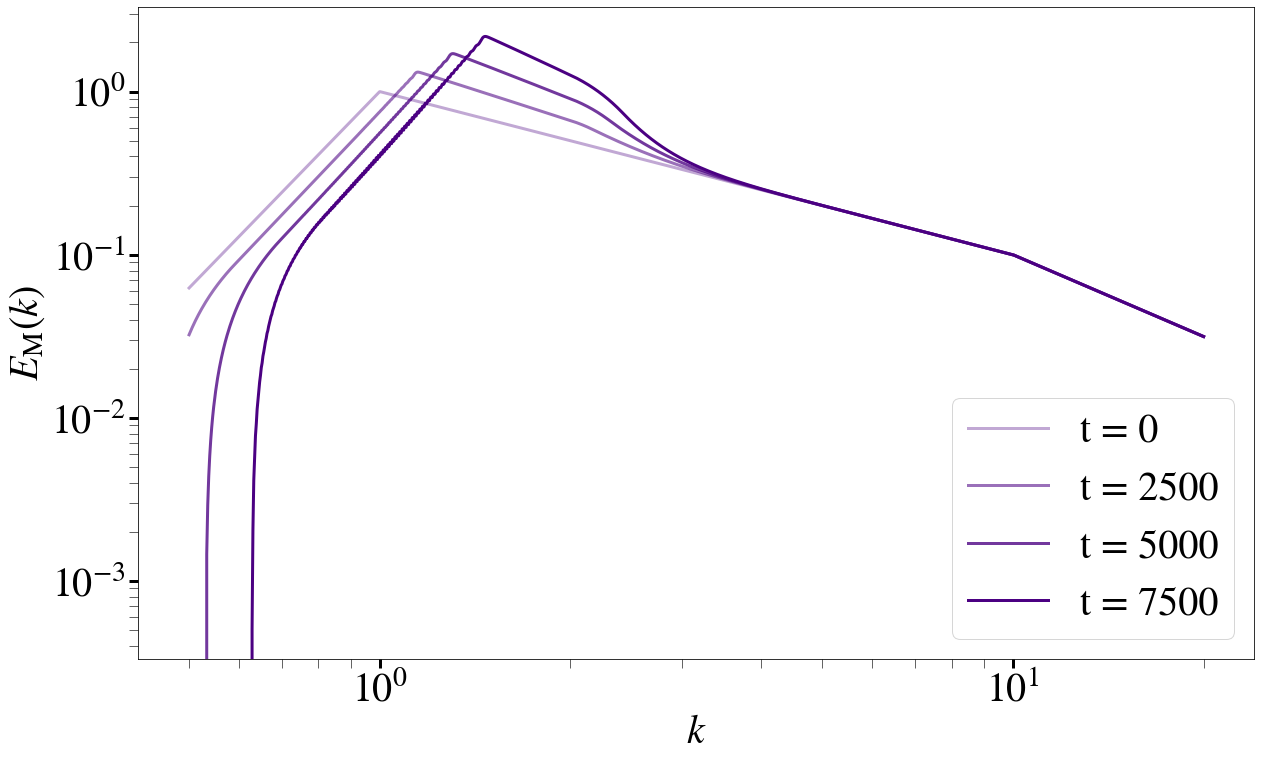}
	\caption{$a=1$}
	\label{FigJeans1}
    \end{subfigure}
    ~ 
    \begin{subfigure}[b]{0.49\textwidth}
	\includegraphics[width=\textwidth]{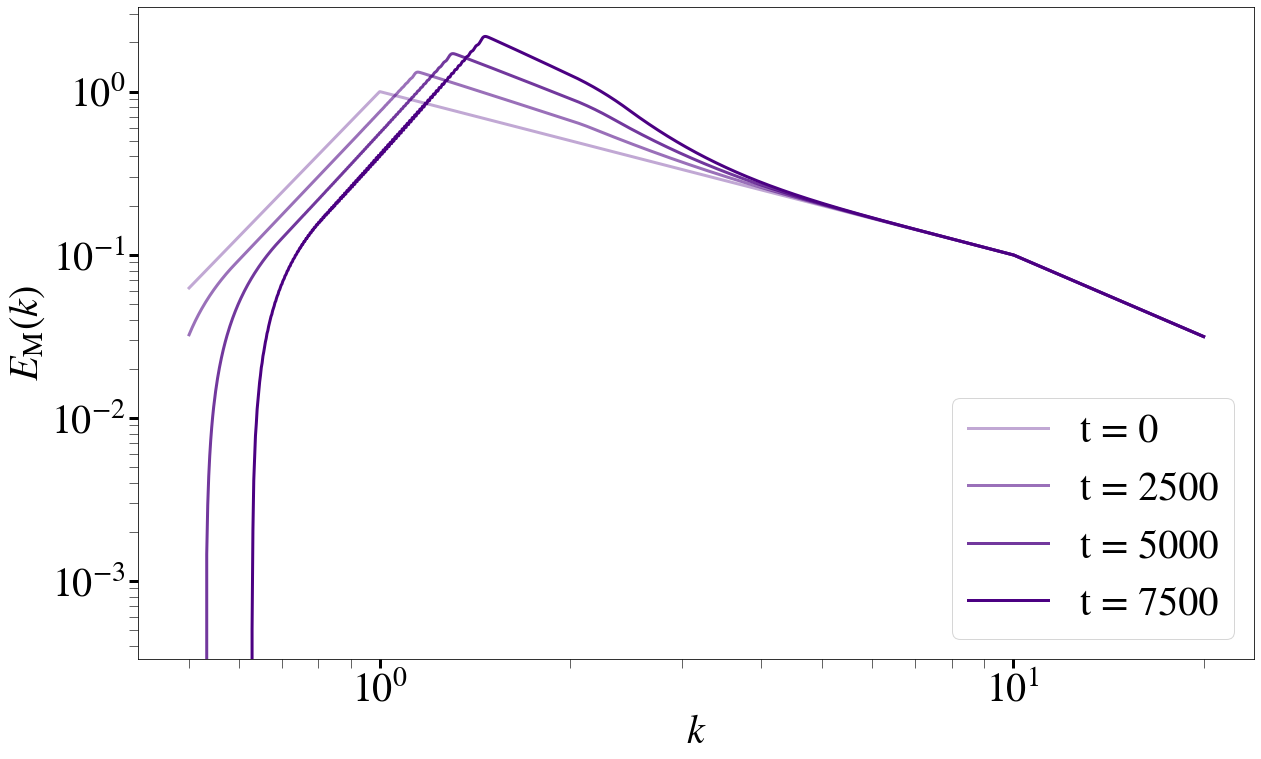}
	\caption{$a=0.5$}
	\label{FigJeans2}
    \end{subfigure}
    \caption{Evolution of the magnetic energy density $E_{\rm M}(k)$ according to Eq.~\eqref{eNewEnDenKInf-3}, and a realistic collapse rate given by Eqs.~\eqref{eZetaSmallK} and \eqref{eJeans}. The time $t$ is measured in units of timesteps. We show the results for two different values of $a$, which characterizes the transition of the collapse rate across the Jeans wavenumber.}
    \label{FigCollapseRateJeans}
\end{figure*}

Then we model the evolution of $E_\mathrm{M}$, using Eq.~(\ref{eNewEnDenKInf-3}).
In Fig.~\ref{FigCollapseRateJeans}, we show the evolution of $E_{\rm M}(k)$ for this realistic collapse rate, using $k_\mathrm{J} = 2$ and $\zeta_0 =0. 03$, for two different values of $a$.
We note here that we choose these values for visual purposes; for other choices of $k_{\rm J}$, $\zeta_0$, and $a$, the solution behaves in an analogous way.
The plots use a time defined as the location along the length grid $N_t = 10000$.
We indeed see a forward collapse for both values of $a$, which is consistent with the expectations.
In contrast to the case of the constant collapse rate $\zeta$, as shown in Fig.~\ref{FigConstRate}, the growth of magnetic energy in the present case is suppressed at small scales (i.e., large $k$) due to the existence of the Jeans scale in the realistic collapse rate $\zeta(k)$, as structures smaller than the Jeans scale do not collapse.
In both cases, the amplification of magnetic fields is the result of magnetic flux compactification due to the gravitational collapse of structures.

\begin{figure*}[!t]
    \begin{subfigure}[b]{0.45\textwidth}
	\includegraphics[width=\textwidth]{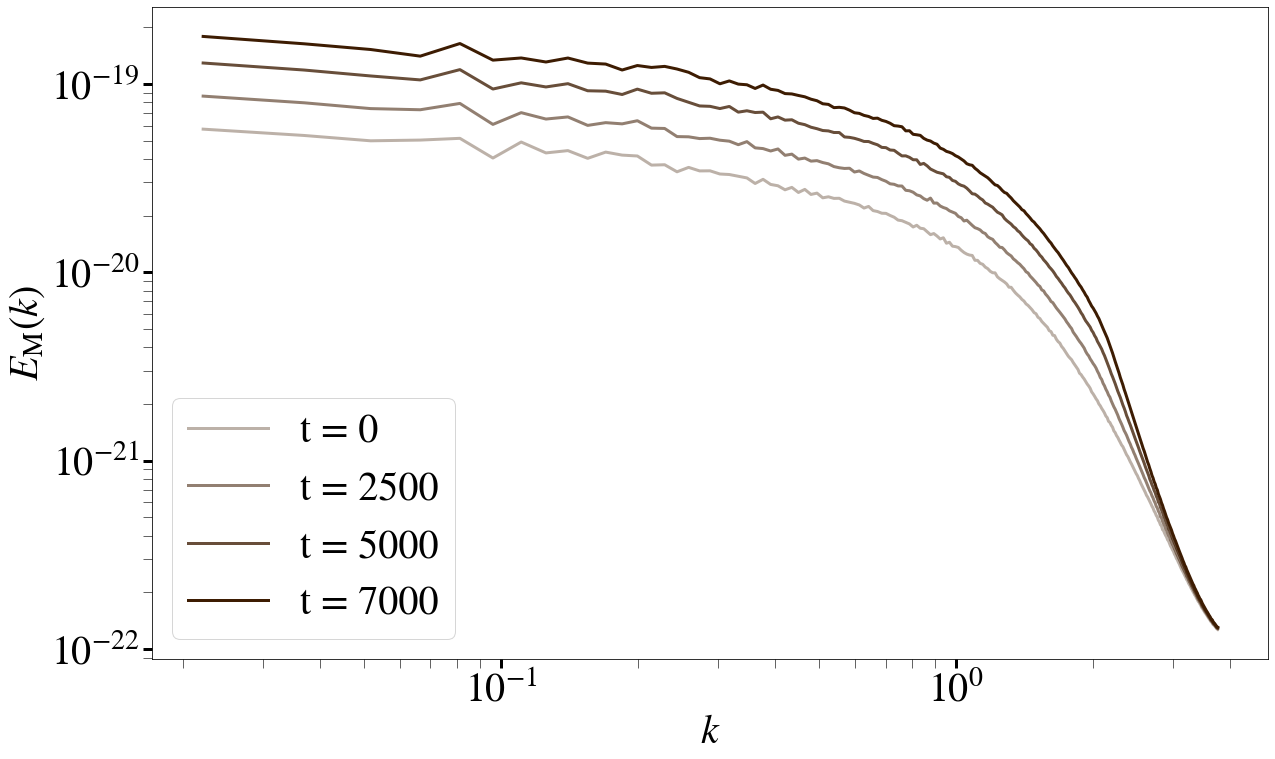}
	\caption{Uniform spectrum.}
	\label{FigUniEvol}
    \end{subfigure}
    ~ 
    \begin{subfigure}[b]{0.45\textwidth}
	\includegraphics[width=\textwidth]{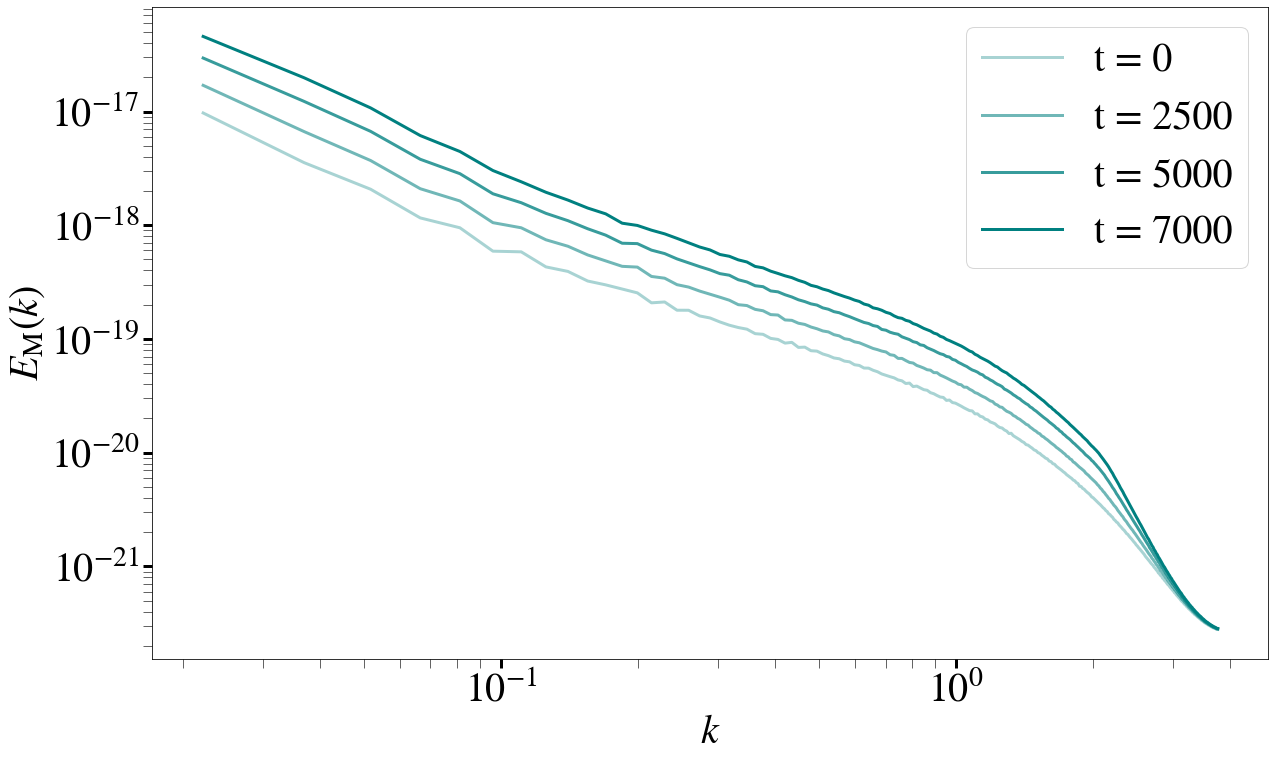}
	\caption{Scale-invariant spectrum.}
	\label{FigSIEvol}
    \end{subfigure}
    
    \begin{subfigure}[b]{0.45\textwidth}
	\includegraphics[width=\textwidth]{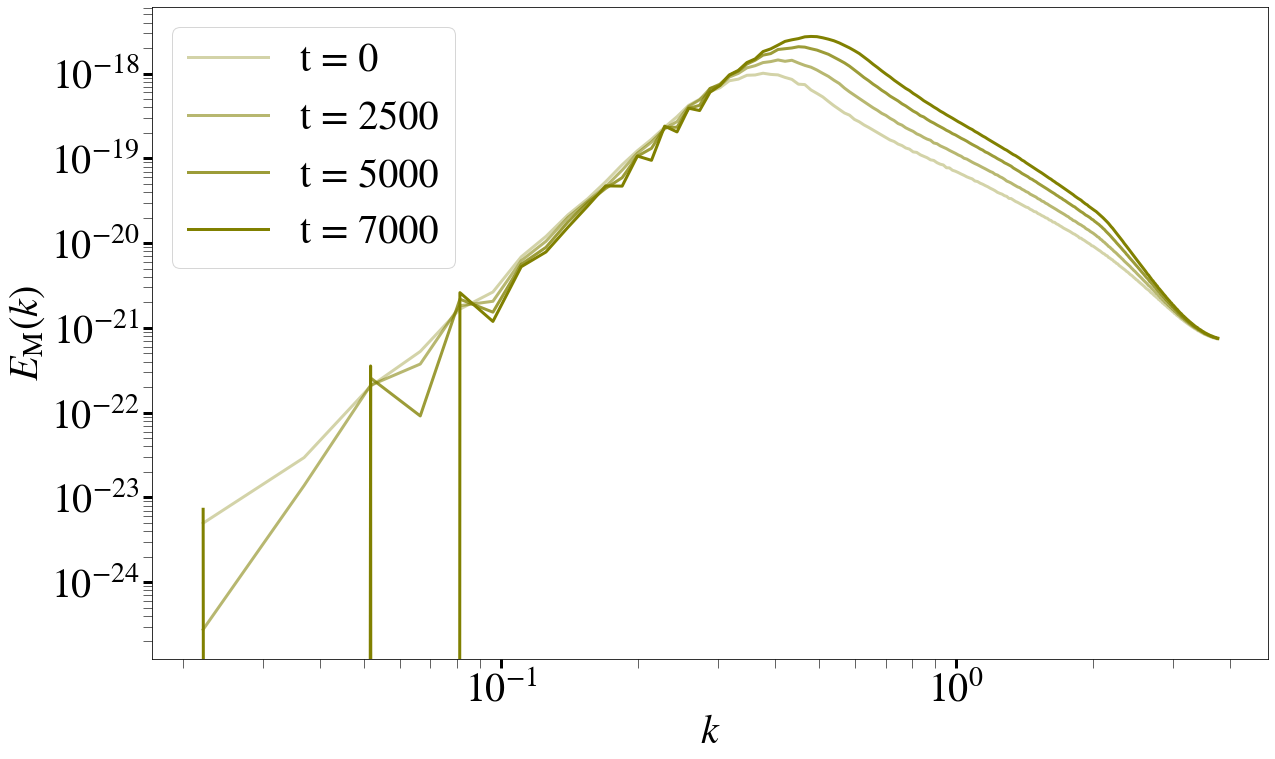}
	\caption{Helical spectrum.}
	\label{FigHelEvol}
    \end{subfigure}
    ~
    \begin{subfigure}[b]{0.45\textwidth}
	\includegraphics[width=\textwidth]{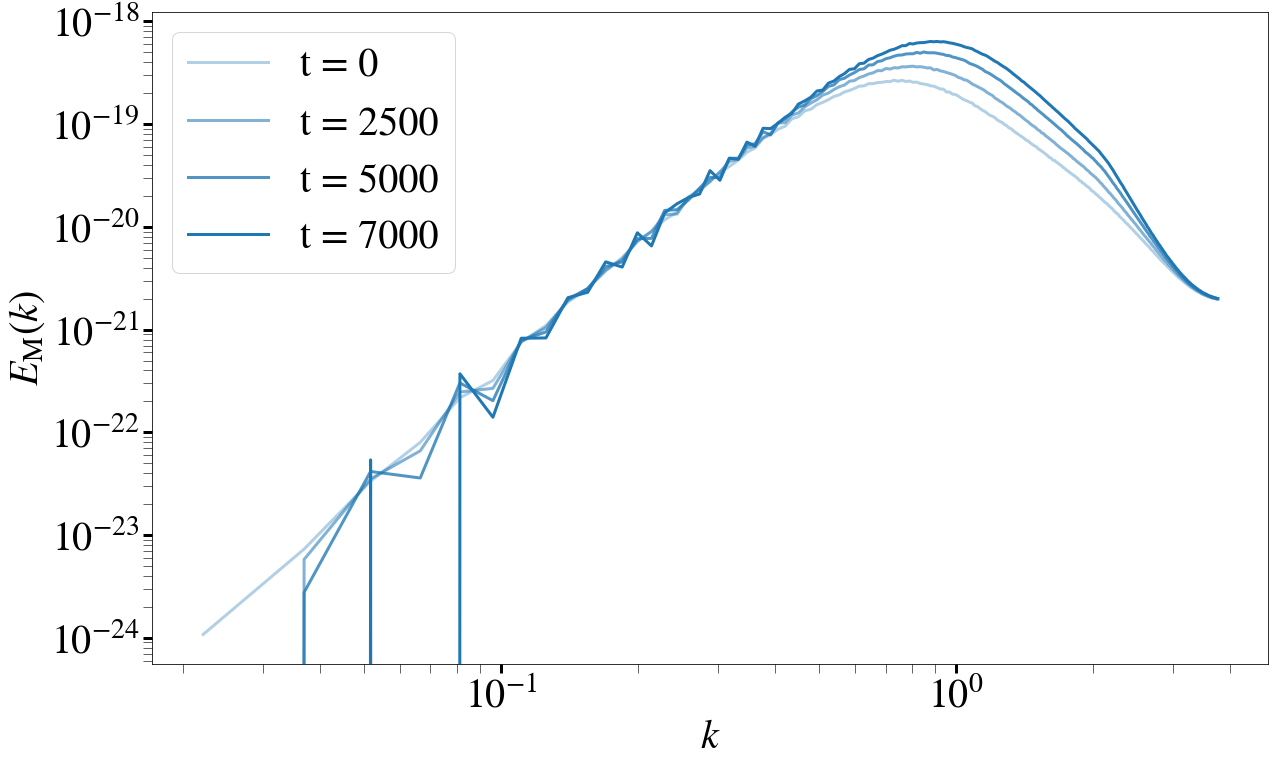}
	\caption{Nonhelical spectrum.}
	\label{FigNonHelEvol}
    \end{subfigure}
    \caption{The evolution of the magnetic energy spectra $E_\mathrm{M}(k)$ with redshift $z$ as affected by structure formation, calculated by numerical simulations in Ref.~\cite{Mtchedlidze:2021bfy}, for different initial spectra for seed PMFs -- uniform, scale-invariant, helical, and 
    nonhelical. There is indication of numerical instability, likely due to the relatively large $\mathrm{d}k$ used in these simulations.}
    \label{FigEvols}
\end{figure*}

We now apply our forward cascade model to the magnetic energy spectra obtained from cosmological simulations in Fig.~\ref{FigEvols}.
Specifically, we use the spectra presented in Fig.~\ref{Fig3D} at redshift $z=10$ as our initial conditions.
We note the similarity between the cosmological simulations and this semi-analytic results, but no visible increase in energy density at small scales in the analytical solution occurs compared to the simulations.
We also note some forward cascade of energy at large scales for the helical and nonhelical cases in the analytical method, which is qualitatively consistent with the cosmological simulations.
The numerical instability is observed for all four spectra, which is a product of a relatively large $\mathrm{d}k \approx 0.079$, due to constraints in the evaluation of the initial conditions.

It should be noted that at small wavenumbers, i.e. those below the peak of the magnetic energy spectra, the magnetic energy slightly increases in the simulations (Fig.~\ref{Fig3D}, panels (c) and (d)), an effect not seen in our analytical model.
This is because in the simulations, the growth of the field in all stochastic cases is dominated by non-linear effects which we think leads to the transfer of energy to both large and small scales.

We see that a forward cascade is absent for large-scale correlated fields, as was also shown by the simulation results (see Figure~\ref{Fig3D}, panels a and b).
Since for the uniform and scale-invariant cases most of magnetic energy is concentrated on large scales, the coupling between small- and large-scale modes is small, and therefore a forward cascade is absent.
We also observe from Fig.~\ref{FigEvols} that the energy uniformly increases on large scales for the uniform and scale-invariant models.
This uniform growth is also seen for the evolution of the uniform model (Fig.~\ref{Fig3D}, panel (a)) in simulations. In simulations, this happens because the amplification of the uniform model is governed by the (linear) growth of density perturbations.
In the scale-invariant case (Fig.~\ref{Fig3D}, panel (b)), on the other hand, from the simulations we see that growth is not uniform and is larger on smaller scales (intermediate and large wavenumbers).

\section{Conclusions}
\label{secConcl}

In this work, we semi-analytically computed the forward cascade of the energy spectrum of primordial magnetic fields in the late stages of the evolution of the Universe.
This evolution is a result of the adiabatic collapse of structures,
which, via the conservation of magnetic flux, strengthens the magnetic fields and pushes them to smaller scales.
We expect that this forward cascade, which occurs during the era of structure formation, partially compensates for the inverse cascade sometimes undergone by the PMFs in the primordial plasma prior to recombination, i.e., at much earlier times\footnote{This inverse cascade depends on many factors, and importantly on the presence of magnetic helicity; see Appendix~\ref{SecMagHelRealiz} for more detail.}.

We showed that this forward cascade depends on the characteristics (i.e., the spectral shape, fractional helicity, etc.) of the magnetic field itself.
For fields that have small coherence scales ($\lesssim 2\Mpc$) we observe forward cascade to be pronounced on scales below the Jeans scale for large-scale correlated fields we do not observe forward cascade.
Similarly to the results from cosmological simulations \cite{Mtchedlidze:2021bfy}, we see that the power spectrum for the uniform model increases homogeneously on large and intermediate scales.
Our analysis, however, does not seem to be applicable on smaller scales where non-linear growth dominates over linear growth; this is true for all different magnetic field models studied in this work.
We must also note that uncertainties in the evolution of magnetic fields at high wavenumbers also exist in simulations, e.g. due to limited resolution~\cite{Vazzaetal2014}.

An accurate understanding of this forward cascade has important implications for understanding and constraining models and mechanisms of the generation and evolution of PMFs.
From an observational standpoint, the forward cascade of magnetic energy implies that the effects of PMFs on LSS would be on smaller scales than in the absence of a forward cascade, and an accurate modeling of this cascade will tell us the relevant scales on which the effects of PMF on LSS are to be studied.
Importantly, since the forward cascade would occur at a much later era compared to the epoch of recombination, these scales would be smaller than the analogous scales on the CMB sky on which PMFs potentially leave their imprint.

In this work, we have considered a simple model of structure collapse, namely a uniform collapse rate of all structures larger than the Jeans length scale.
Such an approximation already highlights the general qualitative features seen in numerical simulations.
There, a forward cascade emerges at the onset of structure formation around $z\sim 10$. 
In the cosmological simulations where the spectral peak lies within the simulation domain, we observe moderate large-scale growth, a decline in magnetic energy near the initial peak, and a gradual transfer of energy to smaller spatial scales. 
This latter behavior is well captured by our analytical toy model.
In order to model the interplay between structures and PMFs more realistically, we have to relax our simple assumption of the form of the collapse rate $\zeta(k)$ we considered in our analysis above, and allow $\zeta$ to depend on other parameters; one such simple extension, where structures on the same scale may collapse at different mass-dependent rates, is discussed in Appendix~\ref{SecForwardCascEvolEqMassDep}.
Such an analysis, which we defer to future work, is expected to capture the effect of variable collapse rate on the forward cascade of magnetic energy, and would be a more realistic comparison to the simulations.

The results of our work open a new direction in the search for the origin and in the study of the amplification of large-scale magnetic fields. 
To better understand how galaxy-cluster-scale fields have been amplified, we need to account for the forward cascade of small-scale correlated fields (possibly originating from phase transitions) during structure formation. 
The forward cascade of magnetic fields not only amplifies such fields on scales of collapsing regions, but also drives turbulence. Therefore, an interplay between field amplification due to forward cascade and turbulence should be carefully studied. This, in turn, may help us understand how efficiently magnetic fields grow during structure formation and how strong (or weak) initial seed magnetic fields must have been.

\vspace{2mm}
{\bf Data availability} --- The source codes used for the
numerical solutions in this study, and data for the different plots, are freely available at \href{https://github.com/mpa710/pmf}{this repository}.

\vspace{2mm}
{\bf Acknowledgments} ---
We acknowledge helpful discussions with Axel Brandenburg, Xiaolong Du, Hoang (Nhan) Luu, Philip Mocz, Franco Vazza, and Mark Vogelsberger.
MA, TK and SMa acknowledge the National Science Foundation (NSF) Astronomy and Astrophysics Research Grants (AAG) Awards AST2307698 and AST2408411;
EC and TK acknowledge the NASA Astrophysics Theory Program (ATP) Award 80NSSC22K0825;
TK and SMt also acknowledge the Shota Rustaveli National Science Foundation (SRNSF) of Georgia, FR24-2606.
SMt also acknowledges the Fondazione CDP (grant no Rif: 2022-2088 CUP J33C22004310003, ``BREAKTHRU'' project).
TK, SMt, and JS are also thankful to the Bernoulli Center in Lausanne for hospitality and the interesting discussions during the program 
``Generation, evolution, and observations of cosmological magnetic fields''
held in April-June, 2024.

\appendix

\section{Magnetic Helicity and Realizability Condition}
\label{SecMagHelRealiz}

Magnetic fields
generated during inflation or PTs may be helical if the underlying generation processes involve parity violating processes (see Refs.~\cite{Caprini:2014mja, Cornwall:1997ms, Field:1998hi, Vachaspati:2001nb, Sigl:2002kt} for a few examples and Ref.~\cite{Kandus:2010nw} for a detailed census of models). Magnetic helicity is an important quantity that impacts the evolution of magnetic fields in a charged plasma.
We define the mean magnetic helicity density over a comoving volume $V$, with fully contained fields lines, as
\begin{equation}\label{eHelDef}
    \mathcal{H}_{\rm M}=\frac{1}{V}\int_{V} 
    \mathrm{d}^3 
    \mathbf{x}\,\mathbf{A}\cdot\mathbf{B},
\end{equation}
where the comoving magnetic vector potential $\mathbf{A}$ is defined such that\footnote{The \textit{physical} field $\mathbf{B}_{\rm phys}$ and vector potential $\mathbf{A}_{\rm phys}$ are given by $\mathbf{B}=a^2\mathbf{B}_{\rm phys}$, and $\mathbf{A}=a\mathbf{A}_{\rm phys}$, and are related as $\mathbf{B}_{\rm phys}=\nabla_\mathbf{y}\times\mathbf{A}_{\rm phys}$, where $\mathbf{y}=a\mathbf{x}$ denotes the physical coordinates.} $\mathbf{B}=\nabla\times\mathbf{A}$.
Magnetic helicity is a nonlocal quantity and can be understood from a topological point of view in terms of the linkage of nonoverlapping magnetic flux tubes~\cite{Brandenburg:2004jv}; in the mathematical literature, this quantity is typically called a \textit{generalized asymptotic form of the Hopf invariant}~\cite{Arnold1974}.
Analogously to Eq.~\eqref{eEnMean}, we can express the mean magnetic helicity density in terms of the spectral magnetic helicity density as
\begin{equation}\label{eHelMean}
    \mathcal{H}_{\rm M}=\int_0^\infty
    \mathrm{d}k 
    \,H_{\rm M}(k).
\end{equation}

Application of the \textit{Cauchy-Buyanovsky-Schwarz inequality} to the distribution of helical  magnetic fields
implies the existence of an upper bound on the magnetic helicity given by the realizability condition\footnote{Alternatively, we can state that there exists a lower bound on $\xi_{\rm M}$ for a given helicity,
\begin{equation}\label{eXiM}
    \xi_{\rm M}^{\rm min}(t)\equiv\frac{|\mathcal{H}_{\rm M}(t)|}{2\mathcal{E}_{\rm M}(t)},
\end{equation}
which implies $\xi_{\rm M}(t)\geq\xi_{\rm M}^{\rm min}(t)$.}
\begin{equation}\label{eRealCond-2}
    |\mathcal{H}_{\rm M}|\leq2 \xi_{\rm M} \mathcal{E}_M;
\end{equation}
we can thus define a quantity called fractional magnetic helicity:
\begin{equation}\label{eFracHel}
\epsilon_M\equiv\frac{\mathcal{H}_M(t)}{2\xi_M(t) \mathcal{E}_M(t)}\in[-1,1].
\end{equation}
During the magnetic field decay the field lines are redistributed in such a way that the fractional helicity reaches its maximal value, i.e. $\epsilon_{\rm M}=1$.
The decay time necessary to obtain the maximal helical configuration is determined by the initial values of $\epsilon_{\rm M}$ (inversely proportional $\epsilon_{\rm M}^2$~\cite{Tevzadze:2012kk}). 

The presence of magnetic helicity strongly affects the evolution of turbulent magnetic fields:
the conservation of magnetic helicity in the highly conductive plasma of the early universe leads to the inverse cascade of energy in the helical magnetic fields, where magnetic energy is transferred from small to larger length scales;
helical magnetic fields also decay more slowly compared to nonhelical fields~\cite{Kahniashvili:2012uj,Copi:2008he,
Brandenburg:2014mwa,Brandenburg:2017neh,Brandenburg:2017rnt,Brandenburg:2018ptt}.
The time evolution of the energy density and the correlation length on dimensional grounds has been studied by several authors.
In particular, if helicity is conserved, the characteristic scale of magnetic fields increases with time, i.e. an inverse cascade of magnetic energy to larger length scales during the MHD evolution occurs~\cite{Kahniashvili:2012uj, Brandenburg:2014mwa, Brandenburg:2017neh}.

For PMFs with a finite integral length scale\footnote{Traditionally, PMFs with energy spectra $E_{\rm M}(k)\propto k^{n_B}$ with ${n_B}<-1$ at low $k$ have a divergent integral scale $\xi_{\rm M}$. However, in cosmological contexts, causality forces the spectrum to have the shape $E_{\rm M}(k)\propto k^4$ for scales outside the Hubble horizon. This ensures that $\xi_{\rm M}$ always converges.}, the realizability condition holds at every $k$, which we now derive.

We start by defining the momentum space analog $\mathbf{\tilde{B}}(\mathbf{k})$ of the real space  magnetic fields,
\begin{equation}\label{eFT1}
\mathbf{B}(\mathbf{x})=\frac{1}{(2\pi)^3}\int \mathrm{d}^3\mathbf{k}\,e^{-i\mathbf{k}\cdot\mathbf{x}}\,\mathbf{\tilde{B}}(\mathbf{k}),
\end{equation}
and express the three-dimensional power spectrum of the magnetic fields $\mathcal{F}_{ij}(\mathbf{k})$ as
\begin{equation}\label{ePowSpecDef}
\left<\tilde{B}_i^*(\mathbf{k})\tilde{B}_j(\mathbf{k}')\right>=(2\pi)^3\delta^3(\mathbf{k}-\mathbf{k}')\mathcal{F}_{ij}(\mathbf{k}).
\end{equation}
In terms of an orthonormal \textit{right-handed} coordinate system $(\mathbf{e}^1,\mathbf{e}^2,\hat{\mathbf{k}})$, we can construct a \textbf{circular polarization basis},
\begin{equation}\label{eCircPolDef}
\mathbf{e}^\pm=\frac{1}{\sqrt{2}}\big(\mathbf{e}^1\pm i\mathbf{e}^2 \big),
\end{equation}
satisfying conditions $\left(\mathbf{e}^+\right)^2=\left(\mathbf{e}^-\right)^2=0$, $\mathbf{e}^+\cdot\mathbf{e}^-=1$, and $\hat{\mathbf{k}}\times\mathbf{e}^\pm=\mp i\mathbf{e}^\pm$.
Since $\mathbf{\tilde{B}}(\mathbf{k})$ is orthogonal to $\mathbf{k}$, we can expand it and its complex conjugate on the basis of circular polarization,
\begin{equation}\label{eCircPolDec}
\begin{aligned}
\mathbf{\tilde{B}}(\mathbf{k})&=\tilde{B}_+(\mathbf{k})\,\mathbf{e}^++\tilde{B}_-(\mathbf{k})\,\mathbf{e}^-,\\
\mathbf{\tilde{B}}^*(\mathbf{k})&=\tilde{B}^*_+(\mathbf{k})\,\mathbf{e}^-+\tilde{B}^*_-(\mathbf{k})\,\mathbf{e}^+.
\end{aligned}
\end{equation}
Then one can compute the symmetric and antisymmetric parts of $\mathcal{F}_{ij}(\mathbf{k})$ in terms of the circular polarization components using Eq.~\eqref{ePowSpecDef} as 
\begin{equation}\label{ePowSpecDecSymAndAntisym}
    \begin{aligned}
    \delta^3(\mathbf{k}-\mathbf{k}')\mathcal{F}_{ii}(\mathbf{k})
    &=\frac{1}{(2\pi)^3}\left[\left<\tilde{B}_+^*(\mathbf{k})\tilde{B}_+(\mathbf{k}')\right>\right.\\
    &\left.+\left<\tilde{B}_-^*(\mathbf{k})\tilde{B}_-(\mathbf{k}')\right>\right],\\
    i\epsilon_{imj}\hat{k}_m\delta^3(\mathbf{k}-\mathbf{k}')\mathcal{F}_{ij}(\mathbf{k})
    &=\frac{1}{(2\pi)^3}\left[\left<\tilde{B}_+^*(\mathbf{k})\tilde{B}_+(\mathbf{k}')\right>\right.\\
    &\left.-\left<\tilde{B}_-^*(\mathbf{k})\tilde{B}_-(\mathbf{k}')\right>\right],
    \end{aligned}
    \end{equation}
where in the steps leading up to these expressions we implicitly set $\mathbf{k}=\mathbf{k}'$ (e.g., while taking the inner product of the basis vectors $\mathbf{e}^\pm$), justifying it due to the presence of the delta function.
In fact, since the LHS of both equations in Eq.~\eqref{ePowSpecDecSymAndAntisym} vanish when $\mathbf{k}\neq\mathbf{k}'$, we can set that in their RHS too, and write
\begin{equation}\label{ePowSpecDecFin}
\begin{aligned}
\left[\left<\left|\tilde{B}_+(\mathbf{k})\right|^2\right>+\left<\left|\tilde{B}_-(\mathbf{k})\right|^2\right>\right]
&=(2\pi)^3\delta^3(\mathbf{k}-\mathbf{k}')\frac{E_{\rm M}(k)}{2\pi k^2},\\
\left[\left<\left|\tilde{B}_+(\mathbf{k})\right|^2\right>-\left<\left|\tilde{B}_-(\mathbf{k})\right|^2\right>\right]
&=(2\pi)^3\delta^3(\mathbf{k}-\mathbf{k}')\frac{H_{\rm M}(k)}{4\pi k},
\end{aligned}
\end{equation}
using the fact that
\begin{equation}\label{eCorFourierEMHM}
	\frac{{\mathcal F}_{ii}({\bf k}) }{(2\pi)^3}=\frac{E_{\rm M}(k)}{2\pi k^2},\quad\quad\frac{i\epsilon_{imj}\hat{k}_m{\mathcal F}_{ij}({\bf k}) }{(2\pi)^3}=\frac{H_{\rm M}(k)}{4\pi k},
\end{equation}
as can be easily checked from Eq.~\eqref{eCorFourier-2}.
We can compare the LHS in the above equations using the fact that $\left|a-b\right|\leq a+b$ for $a,b\in\mathbb{R}^+\cup\{0\}$.

This finally leads to the spectral form of the realizability condition,
\begin{equation}\label{eRealCond-4}
    |H_{\rm M}(k)|\leq 2k^{-1}E_{\rm M}(k).
\end{equation}

\section{Resolution effect for simulations}
\label{App:ResEffect}

\begin{figure}[t]
    \includegraphics[width=\columnwidth]{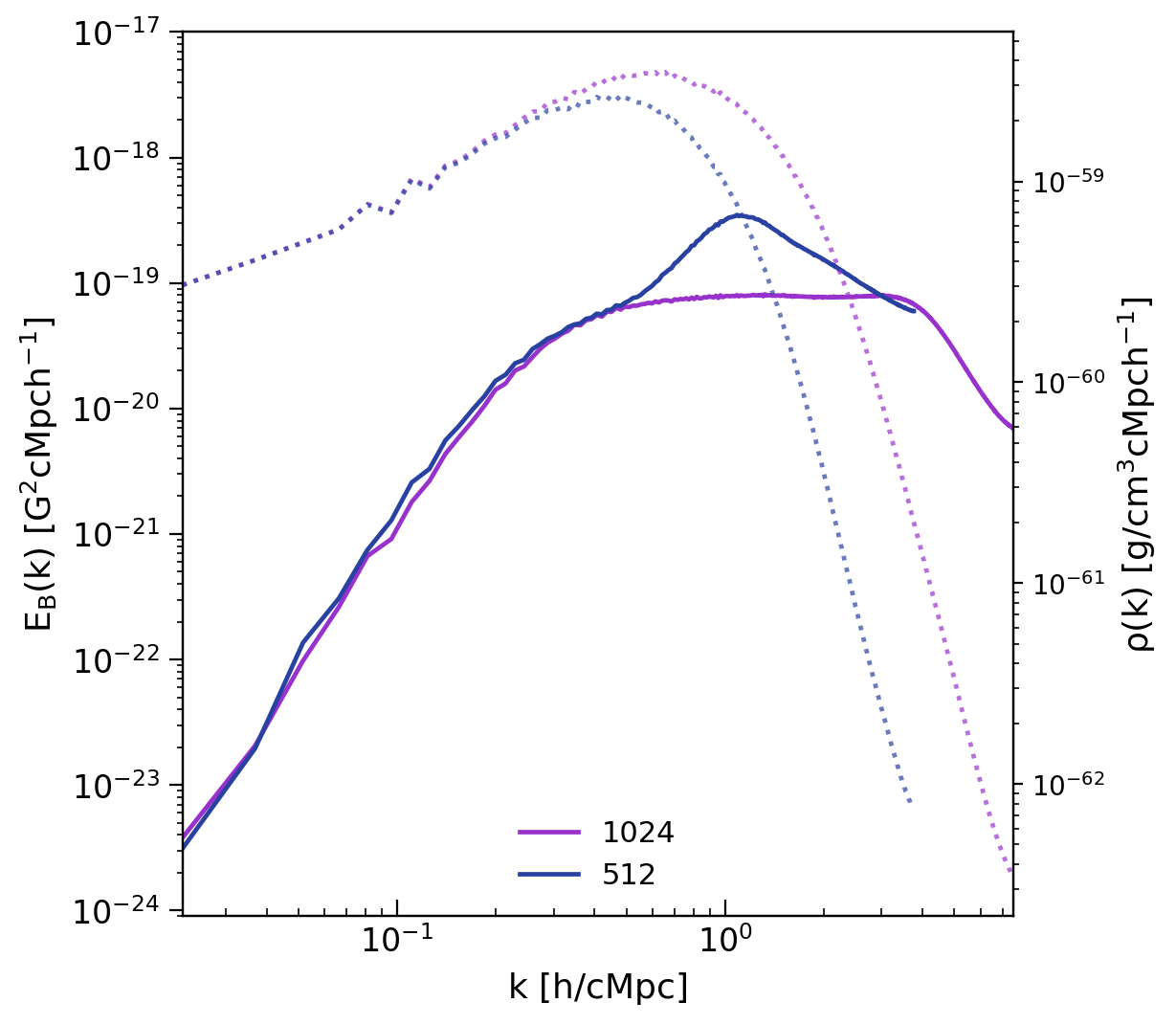}
    \caption{Density (lower-opacity dotted lines) and magnetic energy power spectrum for two different resolutions, for $512^3$ and $1024^3$ grid points, from our cosmological simulations.}
    \label{FigResDep0}
\end{figure}

In Figure~\ref{FigResDep0}  we see that for our higher-resolution  run (with $1024^3$ grid points) the final peak of the magnetic energy is not as pronounced as for the lower-resolution case. However, we still see a less pronounced bump at $k\sim 4 \cMpchI$. This bump is at higher $k$ than the peak for 
the $512^3$ run. This shift to further higher wavenumbers might be governed by the shift of the peak of the density field in the higher-resolution case.

\section{Generalization Considering Mass-Dependent Collapse Rate}
\label{SecForwardCascEvolEqMassDep}

We generalize $\zeta(k,t)$ to define a mass-dependent rate of collapse of the structure $\zeta(m,k,t)$ 
which is the rate at which a mass $m$ on scale $k$ shrinks over time $t$.
We also define $\nu(m,k,t)$ to be the number density of structures of mass $m$ at scale $k$ at time $t$, i.e., $\nu(m,k,t)\,\mathrm{d}m$ is the number of structures at scale $k$ with masses between $m$ and $m+\mathrm{d}m$.
Thus the number of structures at scale $k$ at time $t$ is given by
\begin{equation}\label{eNumStructScale}
    N(k,t)=\int \mathrm{d}m\,\nu(m,k,t).
\end{equation}
We can work out a time evolution of the number density by noting from a generalized form of Eq.~\eqref{eScaleComp-1} that a structure of mass $m$ at time $t+\mathrm{d}t$ at scale $k$ was at a scale $k-\zeta(m,k,t)\,\mathrm{d}t$ at time $t$;
this collapse preserves the number of structures with masses between $m$ and $m+\mathrm{d}m$, i.e.,
\begin{equation}\label{eNumStructScale-2}
    \nu\left(m,k,t+\mathrm{d}t\right)=\nu\left(m,k-\zeta(m,k,t)\,\mathrm{d}t,t\right),
\end{equation}
leading to the evolution equation (suppressing the $(m,k,t)$ dependence of $\zeta$ and $\nu$)
\begin{equation}\label{eNumStructScale-3}
	\frac{\pa\nu}{\pa t}+\zeta\frac{\pa\nu}{\pa k}=0.
\end{equation}
It is convenient to define a fractional density $\pi(m,k,t)$ of structures of mass $m$ at scale $k$ at time $t$ as
\begin{equation}\label{eFracDensDef}
	\pi(m,k,t)\equiv\frac{\nu(m,k,t)}{N(k,t)},
\end{equation}
so that from Eq.~\eqref{eNumStructScale} we get
\begin{equation}\label{eFracDensDef-2}
    \int \mathrm{d}m\,\pi(m,k,t)=1.
\end{equation}

Considering that the rate of structure collapse is mass dependent, we can generalize Eq.~\eqref{eNewEnDenKInf} as
\begin{equation}\label{eNewEnDenKInfMassDep}
    \begin{aligned}
        E_{\rm M}(k,t+\mathrm{d}t)&=\int \mathrm{d}m\,\pi\left(m,k-\zeta(m,k,t)\,\mathrm{d}t,t\right)\\
        &\times\left[1-\frac{\zeta(m,k,t)\,\mathrm{d}t}{k}\right]^{-4}\\
        &\times E_{\rm M}\left(k-\zeta(m,k,t)\,\mathrm{d}t,t\right).
    \end{aligned}
\end{equation}
Again with appropriate binomial and Taylor expansions up to $\mathcal{O}(\mathrm{d}t)$, we can write the above as
\begin{equation}\label{eNewEnDenKInfMassDep-2}
    \begin{aligned}
        &E_{\rm M}(k,t)+\frac{\pa E_{\rm M}(k,t)}{\pa t}\,\mathrm{d}t\\
        &=\int \mathrm{d}m\left\{\left[\pi(m,k,t)-\frac{\pa\pi(m,k,t)}{\pa k}\zeta(m,k,t)\right]\right.\\
        &\left.\times\left[1+\frac{4\zeta(m,k,t)\,\mathrm{d}t}{k}\right]\left[E_{\rm M}(k,t)-\frac{\pa E_{\rm M}(k,t)}{\pa k}\zeta(m,k,t)\,\mathrm{d}t\right]\right\}\\
        &=E_{\rm M}(k,t)\\
        &+\int \mathrm{d}m\,\zeta(m,k,t)\left\{\pi(m,k,t)\left[\frac{4}{k}-\frac{\pa}{\pa k}\right]-\frac{\pa\pi(m,k,t)}{\pa k}\right\}\\
        &\times E_{\rm M}(k,t)\,\mathrm{d}t+\mathcal{O}(\mathrm{d}t^2),
    \end{aligned}
\end{equation}
where in going to the second step, we have used Eq.~\eqref{eFracDensDef-2}.
We can simplify Eq.~\eqref{eNewEnDenKInfMassDep-2} to get the mass-dependent forward cascade equation
\begin{equation}\label{eNewEnDenKInfMassDep-3}
    \begin{aligned}
        &\frac{\pa E_{\rm M}(k,t)}{\pa t}+\int \mathrm{d}m\,\zeta(m,k,t)\\
        &\times\left\{\frac{\pa\pi(m,k,t)}{\pa k}-\pi(m,k,t)\left[\frac{4}{k}-\frac{\pa}{\pa k}\right]\right\}E_{\rm M}(k,t)=0,
    \end{aligned}
\end{equation}
which is a generalized version of the cascade equation Eq.~\eqref{eNewEnDenKInf-3}.

\section{Implicit Numerical Methods and Method of Characteristics}

\begin{figure}[t]
    \includegraphics[width=\columnwidth]{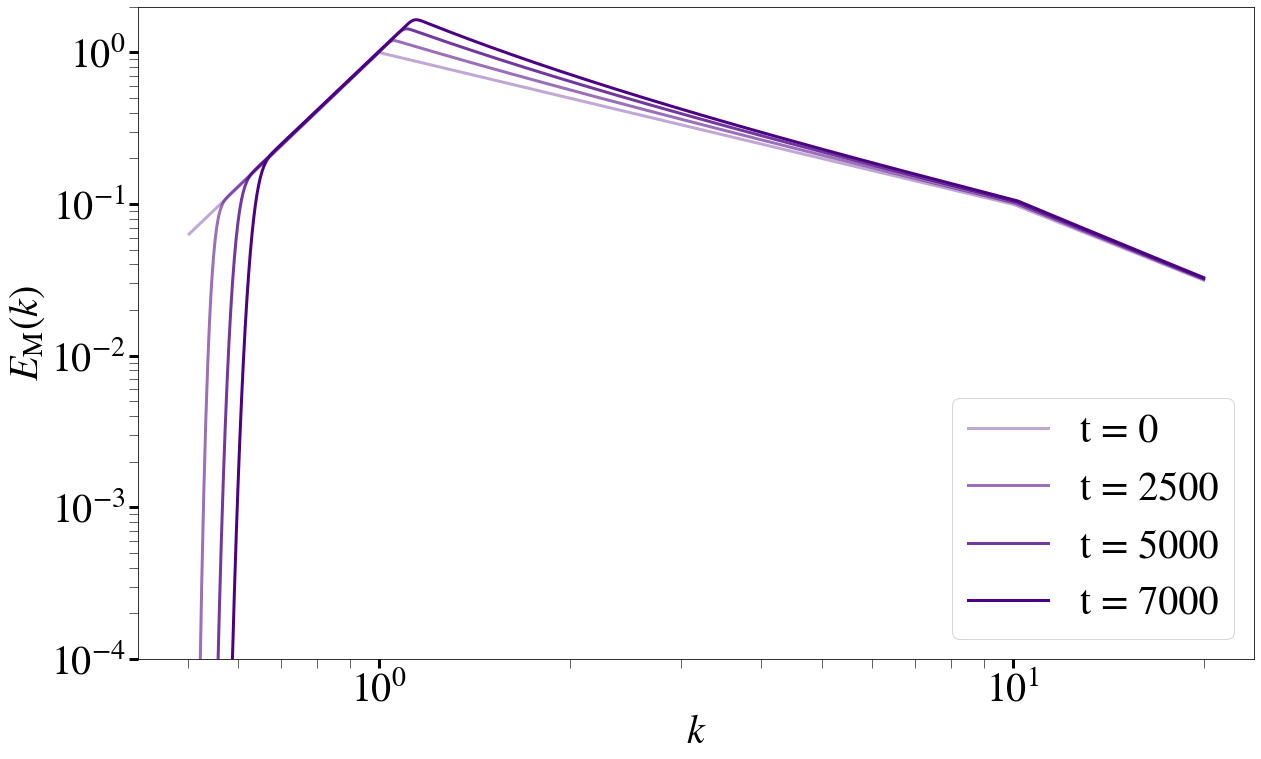}
    \caption{Implicit numerical solution for Eq.~(\ref{eNewEnDenKInf-3}).}   
    \label{FigImplicit}
\end{figure}

\begin{figure}[t]
    \includegraphics[width=\columnwidth]{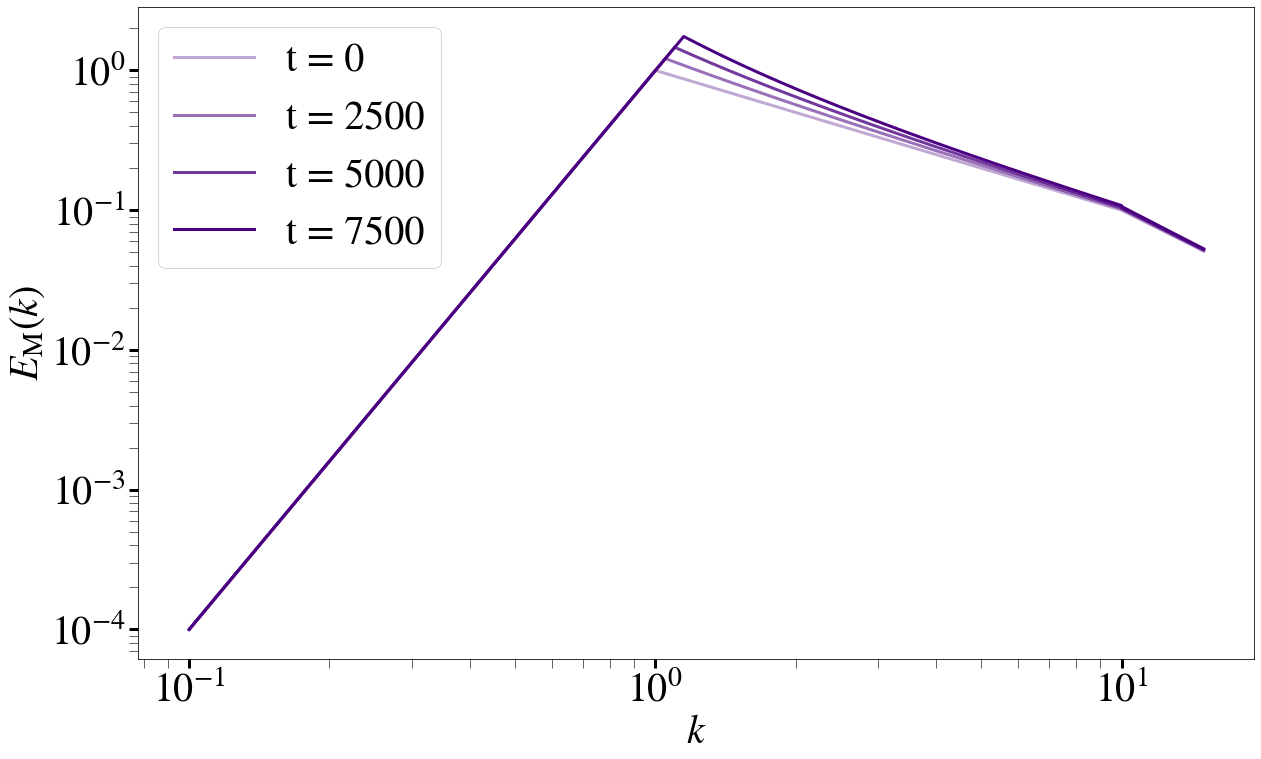}
    \caption{Method of Characteristics solution for Eq.~(\ref{eNewEnDenKInf-3}).
    }
    \label{FigMOC}
\end{figure}

To check the numerical solutions as solved with the Explicit Finite Difference Method, we additionally use the Implicit Finite Difference Method. This method uses the ``inverse" stencil for the explicit method, generating the equation of the form
\begin{equation} 
\frac{E_{\mathrm{M},t+1}^{k} - E_{\mathrm{M},t}^k}{\Delta t} + \zeta_0 \frac{E_{\mathrm{M},t+1}^{k} - E_{\mathrm{M},t+1}^{k-1}}{\Delta k} - \frac{4 \zeta_0}{k} E_{\mathrm{M},t+1}^{k} = 0,
\end{equation}
which yields a system of linear equations for $E_{\mathrm{M},t+1}^{k}$,
\begin{equation}
E_{\mathrm{M},t+1}^{k} \left(1 + \lambda - \frac{4 \lambda \Delta k}{k} \right) - \lambda E_{\mathrm{M},t+1}^{k-1} = E_{\mathrm{M},t}^k.
\end{equation}
Here $\lambda \equiv \zeta_0 \Delta t/\Delta k$ and in matrix form,
\begin{equation}
A = 
\begin{bmatrix}
1 + \lambda - \frac{4 \lambda \Delta k}{k} & 0 & ...\\ 
-\lambda & 1 + \lambda - \frac{4 \lambda \Delta k}{k} & ... \\
...& -\lambda &  1 + \lambda - \frac{4 \lambda \Delta k}{k} 
\end{bmatrix};
\end{equation}
this yields 
\begin{equation}
A E_{\mathrm{M},t+1}^{k} = E_{\mathrm{M},t}^k,
\end{equation}
and the solution of this system of linear equations is shown in Fig.~\ref{FigImplicit}.

The method of characteristics solves for the evolution of the magnetic energy density with a fixed rate of collapse $\zeta_0$.
We start with the Lagrange-Charpit equations, 
\begin{equation}\label{eLagChar}
\frac{\mathrm{d}t}{1} = \frac{\mathrm{d}k}{\zeta_0} = \frac{k\,\mathrm{d}E_{\rm M}}{4 \zeta_0E_{\rm M}}.
\end{equation}
Substituting
\begin{equation} 
k(t) = \zeta_0t+k_0
\end{equation}
into Eq.~\eqref{eLagChar} and solving, we get
\begin{equation}
\ln(E_{\rm M}) = 4\ln(k(t))+C' \implies E_{\rm M}= Ck^4.
\end{equation}
The initial conditions imply that
\begin{equation}
    C = \frac{E_0(k_0)}{k_0^4},
\end{equation}
where $E_0(k)$ is the initial condition function, solved at $t=0$.
We therefore arrive at the final equation,
\begin{equation}\label{eMOC}
    E_{\rm M}(k,t) = E_0(k-\zeta t)\left [ \frac{k}{k-\zeta_0 t} \right]^4,
\end{equation}
which is plotted in Fig.~\ref{FigMOC}.

\section{Comparison of Free Fall Time in Numerical Simulations to Classical Free Fall Time}

To make a crude estimate of the collapse of this material, we start with the known free fall time for a pressure-less, uniform sphere of free fall time.

\begin{equation}\label{etFF}
    t = \sqrt{\frac{3 \pi}{32 G \rho_{\rm{max}}}},
\end{equation}

and also write the Newtonian force equation for an arbitrary particle of mass $m$ on the surface of the sphere with mass $M$, where $r(t)$ is taken as the radius of the sphere, as

\begin{equation}\label{eRadShrink-1}
    \ddot{r}(t) = - \frac{G M}{r(t)^2}.
\end{equation}

The mass $M$ can be rewritten in terms of density and radius as $\frac{4}{3} \pi r^3(t) \rho_{\mathrm{max}}$, and we assume that there is a uniform density $\rho_ = \rho_{\mathrm{max}}$ throughout the space occupied by the material. 

\begin{figure}[t]
\centering
\includegraphics[width=\columnwidth]{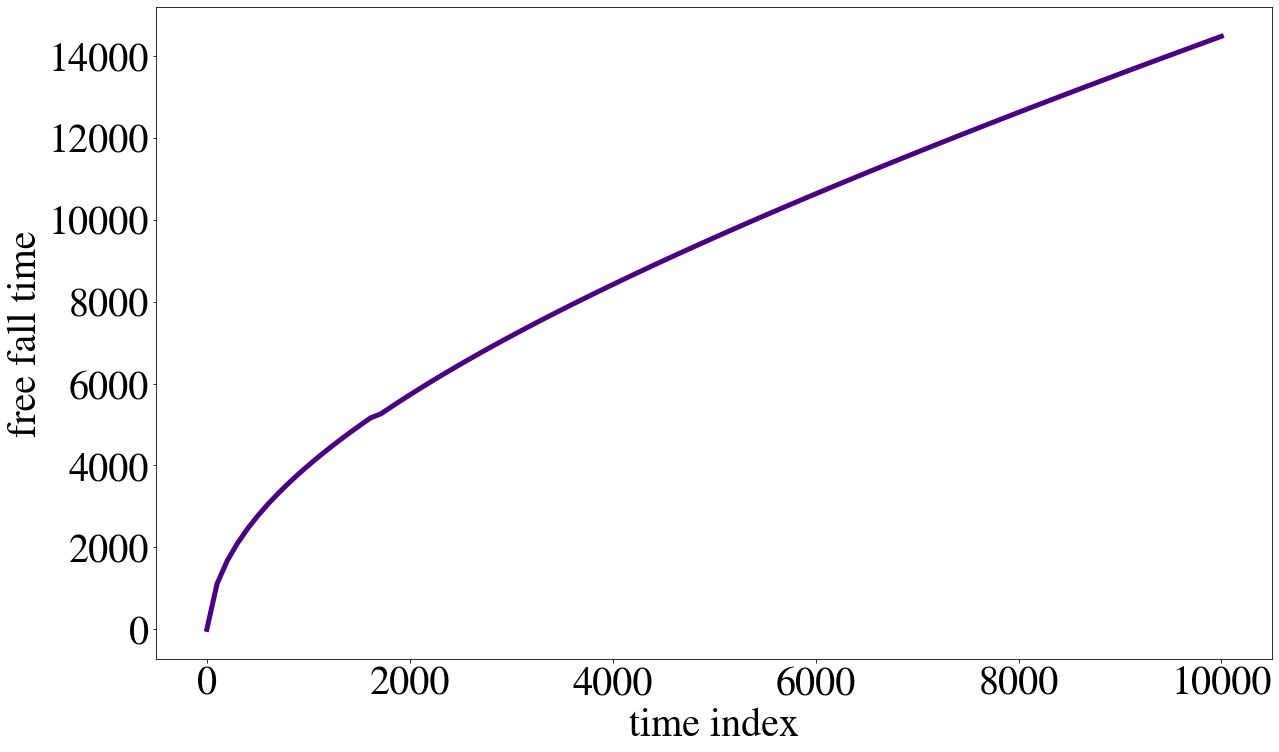}
\caption{Free Fall Time Versus Time Index for Fig.~\ref{FigConstRate}, using Eq.~(\ref{EqFreeFallCollapseForm}).
}
\label{FigFFT}
\end{figure}

Substituting the above definition of $M$ in Eq.~\eqref{eRadShrink-1}, we obtain

\begin{equation}\label{eRadShrink-2}
    \ddot{r}(t) = - \frac{4 \pi \rho_{\mathrm{max}}G r(t)} {3}.
\end{equation}

Defining $\omega^2 = 4 \pi \rho_{\rm{max} }G/3$, we see that our solution is

\begin{equation}\label{eRadCollSol-1}
    r(t) = r_0 \cos(\omega t),
\end{equation}

where $r_0$ is the radius at $t = 0$, and the natural sinusoidal term vanishes due to the initial collapse rate of the sphere being assumed to be zero.
Under the assumption that, in our solutions, $\lambda \sim r \implies \frac{2 \pi}{k} \sim r$, and taking $r_0 \sim k_0 $ , where $k_0$ is unity in our simulations, we rewrite this equation as follows.

\begin{equation}\label{EqFreeFallCollapseForm}
    \arccos\left(\frac{1}{k}\right) = \omega t = \frac{\pi t}{2\sqrt{2} t_{\rm{ff}}},
\end{equation}

where $k$ is the location of the peak in Fig.~\ref{FigConstRate} for a given time $t$, where the time index versus the time of free fall is plotted in Fig.~\ref{FigFFT}. We note an increase in free fall time with respect to time, an error likely attributable to the assumption of fixed density throughout space.

\bibliographystyle{apsrev4-2}
\bibliography{PMF-LSS}

\end{document}